\documentclass[12pt]{article}
\usepackage{epsfig}
\usepackage{a4}

\def\to{\rightarrow}

\def\ra{\rightarrow}
\newcommand\nutau{{\nu_\tau}}

\newcommand\numu{{\nu_\mu}}
\newcommand\anumu{\bar{\nu}_\mu}
\newcommand\nue{{\nu_e}}
\newcommand\anue{\bar{\nu}_e}
\def\sq2{sin^2(2\Theta)}

\def\NOE{{\em N\raise.5ex\hbox{O \kern-0.47em}E\kern.4em}}

\def\dms2{\Delta m^2}
\def\Dm32{\Delta m^2_{32}}
\def\Dm12{\Delta m_{12}^2}
\def\tet12{\theta_{12}}
\def\tet23{\theta_{23}}
\def\tetonethree{\theta_{13}}

\def\Journal#1#2#3#4{{#1} {\bf #2} (#3) #4.}
\def\etal{{\it et\ al.}}

\def\NIM{\em Nucl. Instrum. Methods}

\def\NPB{{\em Nucl. Phys.} B}
\def\PLB{{\em Phys. Lett.}  B}
\def\PRL{\em Phys. Rev. Lett.}
\def\PRC{{\em Phys. Rev.} C}
\def\PRD{{\em Phys. Rev.} D}
\def\ZPC{{\em Z. Phys.} C}

\begin{document}

\thispagestyle{empty}
\begin{flushright}
{\tt ICARUS/TM-2000/01}\\ 
\today
\end{flushright}
\vspace*{1cm}
\begin{center}
{\Large{\bf Physics potential at a neutrino factory:
can we benefit from more than just detecting muons?} }\\
\vspace{.5cm}
A. Bueno\footnote{Antonio.Bueno@cern.ch},
M. Campanelli\footnote{Mario.Campanelli@cern.ch}
and A. Rubbia\footnote{Andre.Rubbia@cern.ch}

\vspace*{0.3cm}
Institut f\"{u}r Teilchenphysik, ETHZ, CH-8093 Z\"{u}rich,
Switzerland

\end{center}
\vspace{2.cm}
\begin{abstract}
\noindent
In order to fully address the oscillation processes at a neutrino
factory, a detector should
be capable of identifying  and measuring all three charged lepton flavors
produced in charged current interactions {\it and} of measuring 
their charges to discriminate the incoming neutrino helicity. 
This is an experimentally
challenging
task, given the required detector mass for long-baseline experiments. 
We address the benefit of a high-granularity,
excellent-calorimetry non-magnetized target-detector, which
provides a background-free identification of electron neutrino charged current 
and a kinematical selection of tau neutrino charged current
interactions.
We assume that charge discrimination is only available
for muons reaching an external magnetized-Fe spectrometer.
This allows the clean
classification of events into electron, right-sign muon, wrong-sign
muon and no-lepton categories.
In addition, high granularity permits a clean detection
of quasi-elastic events, which by detecting the final state
proton, provide a
selection
of the neutrino electron helicity without the need of an electron charge
measurement. 
From quantitative analyses of neutrino oscillation scenarios, we
conclude that in many cases the discovery sensitivities and the measurements of
the oscillation parameters
are dominated by the ability to measure the muon charge.
However, we identify cases where identification of electron and tau samples 
contributes significantly.
\end{abstract}

\newpage
\pagestyle{plain} 
\setcounter{page}{1}
\setcounter{footnote}{0}

%

\section{Introduction}

The firmly established disappearance of muon neutrinos of cosmic ray 
origin~\cite{superk} strongly points toward the existence of neutrino
oscillations~\cite{pontecorvo}. 
The first generation long baseline (LBL) experiments ---
K2K~\cite{k2k}, MINOS~\cite{minos}, OPERA~\cite{opera} and
ICANOE~\cite{icanoe} --- 
will give a conclusive and unambiguous signature of the oscillation
mechanism 
and will provide the first precise measurements of the parameters governing the 
oscillation mechanism. MiniBOONE~\cite{boone} and the LBL programs will
test the LSND signal\cite{LSND}.

A neutrino ``factory''~\cite{geers,nufacwww} is based on the 
decay of muons circulating in a storage ring.
Neutrino factories raised the interest of
the physics community, since they appear natural follow-ups to the current 
experimental LBL program and could
open the way to future muon colliders. 

As many studies have
shown~\cite{us1,Bueno:1998xy,rujula,barger}, the
physics potential of such facilities are indeed very vast.
An entry-level neutrino factory could
test the LSND signal in a background free environment\cite{Bueno:1998xy}.
More importantly, a neutrino factory source would be of
sufficiently high intensity to perform very long baseline
(transcontinental) experiments. 
It could also bring the neutrino sector into the realm of precision 
measurements. 

The neutrino oscillation phenomenology 
may be complicated and involve
a combination of transitions to $\nue$, $\numu$ and $\nutau$. 
It is quite evident that
future neutrino factories will provide ideal conditions
for the neutrino oscillation physics\cite{us1,rujula,barger,us2,cp}. 
The neutrino flavor phenomenology
could be completely explored:
a precise measurement of the mass difference and mixing matrix
elements is achievable, a test of the unitarity of the mixing
matrix can be performed,
a direct detection of Earth matter effects is feasible~\cite{us2} and
CP violation effects could be studied on the 
leptonic sector~\cite{cp}.

The combination of data from atmospheric neutrino and first generation 
LBL experiments will provide some preliminary information on the
possible sub-leading electron mixing\cite{icanoe}. 
A neutrino factory can largely improve the sensitivity on this 
mixing angle.

Neutrino sources from muon decays provide clear advantages over
neutrino beams from pion decays.
The exact neutrino helicity composition is a fundamental tool to study
neutrino oscillations. It can be easily selected,
since $\mu^+\ra e^+\nue\bar\numu$
and $\mu^-\ra e^-\bar\nue\numu$ can be separately obtained. 

At a neutrino factory,
one could independently study the following flavor transitions:
\begin{eqnarray}
\mu^-\ra e^-& \bar\nue&\numu \nonumber \\ 
& & \ra \nue\ra e^-\rm \ appearance\\
& & \ra \numu\rm \ disappearance, \ same \ sign \ muons\\
& & \ra \nutau\ra\tau^-\rm \ appearance, \ high \ energy \ nu's\\
 & \ra &\bar\nue\rm \ disappearance\\
 & \ra &\bar\numu\ra\mu^+\rm \ appearance, \ wrong \ sign \ muons\\
 & \ra &\bar\nutau\ra\tau^+\rm \ appearance, \ high \ energy \ nu's
\end{eqnarray}
plus 6 other charge conjugate processes initiated
from $\mu^+$ decays.

The other main advantages over traditional pion beams
are (1) the beam is free of systematics
and the composition is well known, therefore ideal for
disappearance studies; (2) the two neutrinos in the beam have
opposite helicities, therefore one can envisage oscillation appearance
searches without intrinsic beam backgrounds (3) the muon
energy is monochromatic and in principle adaptable (4) muon storage rings
allow for multiple baselines, and hence a complete
exploration of the $L/E$ domains of oscillations
(5) because very high intensity will be needed for muon colliders,
very intense muon sources will produce
very intense neutrino sources, at least a factor 100 more
intense than existing high energy facilities.

While physics motivations are well understood, it is not
yet clear which design of
detector would best allow to take full advantage
of the neutrino factory beams.

We think that, in order to fully explore 
the neutrino oscillation processes,
the detector should be capable of:
\begin{enumerate}
\item measuring and identifying all three lepton flavors: 
electron, muon and tau;
\item measuring the sign of the lepton charge;
\item separating between charged and neutral current
interactions.
\end{enumerate}
Experimentally, it
is a very challenging task to build detectors with (1) mass
scales of the order of tens of ktons required for long-baseline
experiments,
(2) with sufficient granularity to cleanly identify
electron and tau leptons and (3) which measure the charge
of these leptons.

Various solutions have been explored recently. One based on nuclear
emulsions and magnetized iron has been discussed
in~\cite{Harris:2000yg}. The main challenge there is to
reach the required mass. Large magnetized 
calorimeters have been discussed in~\cite{cervera}. Such high-density
detectors,
while ``easily'' conceived as massive objects,
have intrinsically very coarse granularity and only allow the clean
measurement of muons. They certainly do not have sufficient power to 
adequately identify
and measure electron or tau charged current states.

In this paper, we are motivated by the recent progress made in 
the direction of the design of the multikton ICANOE detector\cite{icanoe}: 
accordingly, we consider a 10~kton (fiducial) 
high granularity low density liquid argon imaging 
target, complemented with 
a high-acceptance external muon spectrometer. 

Thanks to its extremely high granularity target and its excellent
calorimetric properties, 
this design provides the clean identification and
measurement of all three neutrino flavors: electron, muons and taus. However,
only the sign of the muons reaching the muon spectrometer
can be determined\footnote{The possibility of the measurement of the electron
charge will be addressed in a future work.}.

The aim of this paper is to understand the potentials of 
a non-magnetized high-granularity target detector which,
compared to traditional
high density iron calorimeters,  brings 
the measurement of electrons and taus.

We study the physics potentials of such a detector configuration
for three-family neutrino mixing and
for three different baselines (732, 2900
and 7400 km). 

In Section~\ref{sec:neuosc} we summarize three-family neutrino mixing
framework, including the treatment of propagation through matter.

In Section~\ref{sec:evratedist}, we explain in details how the event
distributions and rates are obtained from a detailed
simulation of neutrino interactions and detector effects.

In Section~\ref{sec:evtclass}, we construct
for given oscillation scenarios, event variable
distributions for various event classes. All events
can be subdivided into the electron,
the same sign muons, the wrong sign muons and the
no lepton samples. 

In addition, in Section~\ref{sec:kineanal} we discuss
the possibility to further discriminate final states
between $\nue$ and $\numu$ origins from $\nutau$ by
means of kinematical analysis of the events.

We also address in Section~\ref{sec:quasi}
the possibility to tag quasi-elastic events which provide
indirect neutrino helicity discrimination however
at a large statistical price.

Section~\ref{sec:fit} is devoted to describing our fits
of the oscillation parameters. The fits are expected to
give back the input reference oscillation parameters and
are used to estimate the precision with which we can estimate
these parameters.

Section~\ref{sec:mixing} presents the results for the important
case in which the oscillation effects can be approximated by
one mass scale.

Since the information about the oscillation parameter is redundantly available
in the visible energy distributions of the various event classes,
we address in Section~\ref{sec:taus} the question of the consistency between the
different observed oscillations processes. The ability to treat the
appearance of electron or tau neutrinos gives good over-constraints on
the mixing matrix.

Finally, in Section~\ref{sec:cp}, we analyze the three-family
scenario including possible $CP$-violation.

\section{Three-family neutrino oscillation framework}
\label{sec:neuosc}

\subsection{Mixing matrix parameterization}
We consider neutrino oscillations in a three-family scenario:
the flavor eigenstates
$\nu_\alpha(\alpha= e,\mu,\tau)$ are related to the mass eigenstates
$\nu'_i(i=1,2,3)$ by the mixing matrix U
\begin{equation}
\nu_\alpha=U_{\alpha i}\nu'_i
\end{equation}
and we parameterize it as:
\begin{equation}
U=\left(
\begin{tabular}{ccc}
$c_{12}c_{13}$      & $s_{12}c_{13}$   &  $s_{13}e^{-i\delta}$ \\
$-s_{12}c_{23}-c_{12}s_{13}s_{23}e^{i\delta}$ &
$c_{12}c_{23}-s_{12}s_{13}s_{23}e^{i\delta}$ & $c_{13}s_{23}$ \\
$s_{12}s_{23}-c_{12}s_{13}c_{23}e^{i\delta}$ &
$-c_{12}s_{23}-s_{12}s_{13}c_{23}e^{i\delta}$ & $c_{13}c_{23}$ 
\end{tabular}\right)
\end{equation}
with $s_{ij}=\sin\theta_{ij}$ and $c_{ij}=\cos\theta_{ij}$. 
We confine without loss of generality 
the mixing angle $\tetonethree$ to values in the
interval $[0, \pi/4]$, and $\theta_{12}$, $\tet23$,
$\delta$ to the interval $[0,\pi/2]$. We present 
results in terms of
$\sin^22\theta_{13}$, $\sin^2\theta_{23}$ and $\sin^2\theta_{12}$,
all running in the interval $[0,1]$. The reason for this
choice can be for example seen in Appendix~A, where we recall oscillation
probabilities in the one mass scale approximation. The oscillation
probabilities for $\nue\ra\numu$ and $\nue\ra\nutau$ depend
on $\sin^22\theta_{13}$ and on the sin and cosine of $\theta_{23}$.
The angle $\theta_{23}$ must span the interval $[0,\pi/2]$, however,
the $\tetonethree$ can 
vary within interval $[0, \pi/4]$. Note that we consider small $\tetonethree$ angles,
so the $\cos^4\tetonethree$ dependence in $\numu\ra\nutau$ is very mild.

For $\delta=0$ (i.e. $U$ is real),
the general expression for the three-family neutrino oscillation probability
is:
\begin{equation}\label{eq:oscill}
P(\nu_\alpha\ra\nu_\beta;E,L) =  P(\bar\nu_\alpha\ra\bar\nu_\beta;E,L)= 
\delta_{\alpha\beta}-4\sum_{j>k} J_{\alpha\beta j k}\sin^2\left(
\Delta_{jk}\right)
\end{equation}
where in natural units $\Delta_{jk}\equiv\Delta m^2_{jk}L/4E=
(m^2_j-m^2_k)L/4E$, $E$ is the neutrino energy, $L$ is
the neutrino path-length, and the Jarlskog term is
$J_{\alpha\beta j k} = U_{\beta j} U_{\beta k} U_{\alpha j}
U_{\alpha k}$. 

We naturally assign the mass difference squared $\Delta m^2_{12}$ to
explain the solar neutrino deficit and the mass difference squared $\Delta
m^2_{32}$ to describe the atmospheric neutrino anomaly. We will take
as reference value $\Delta m^2_{32}=3.5\times 10^{-3}\ \rm eV^2$. 
To cover possible ranges of this value, we will also consider
two other values $\Delta m^2_{32}=5\times 10^{-3}\ \rm eV^2$
and $\Delta m^2_{32}=7\times 10^{-3}\ \rm eV^2$. We will always
assume maximal (2-3)-mixing $\sin^2 2\theta_{23}=1$.
In case of the solar neutrino deficit solution, 
the values for $\Delta m^2_{12}$ and mixing $\sin^2 2\theta_{12}$ 
are not uniquely defined by experiments. We will limit
ourself to the LMA-MSW solution with parameters
$\Delta m^2_{12}=1\times 10^{-4}\ \rm eV^2$ and hypothesize
a maximal (1-2)-mixing $\sin^2 \theta_{12}=0.5$.

In this paper,
we do not consider the LSND result which would force us to include more
states with new parameters beyond three-family mixing.

This choice implies that we will always work in a situation where
$|\Delta m^2_{21}|<|\Delta m^2_{32}|\approx |\Delta m^2_{31}|$.
In first approximation, 
the oscillation phenomena governed by the two mass differences decouple and
the effects produced by $\Delta m^2_{12}$ are small at high energy
for the considered baselines. In the first part of this paper, we will
neglect $\Delta m^2_{12}$ effects and work in the so-called ``one
mass scale approximation''\cite{lipari}. In a second phase, we will
be concerned with $CP$-violation effects and will have to
include $\Delta m^2_{12}$.

For simplicity, we will take $m_1 < m_2 <m_3$
which implies $\Delta m^2_{32}>0$. We recall that neutrino oscillations
through matter can
be used to distinguish $\Delta m^2_{32}>0$ from $\Delta m^2_{32}<0$.

In vacuum, we will express the oscillation probability as a function of
the seven following parameters:
(a) the three mixing angles $\theta_{12}$,$\theta_{13}$,$\theta_{23}$;
(b) the two mass differences squared $\Delta m^2_{12}$, $\Delta m^2_{32}$;
(c) the baseline L;
(d) the neutrino energy $E$.

\subsection{Matter effects}
Since we will consider very long distances between neutrino production and
detection, this will only be possible in practice for neutrinos traveling
inside the Earth. In this case, the neutrino oscillation
probabilities will be modified by an additional diagram due to the 
interaction of electron neutrino with the electrons in the
matter\cite{msw}.
One can maintain the neutrino oscillation formalism derived in
vacuum but define effective masses and mixing angles valid in matter.
For example, the effective masses will result from the diagonalization
of the Hamiltonian:
\begin{equation}
 U \left( 
\begin{array}{ccc}
m_1^2 & 0 & 0 \\
0 & m_2^2 & 0 \\
0 & 0 & m_3^2 \\
\end{array}
\right)U^\dag + \left( 
\begin{array}{ccc}
D & 0 & 0 \\
0 & 0 & 0 \\
0 & 0 & 0 \\
\end{array}
\right)
\end{equation}
where
\begin{equation}
D=2\sqrt{2}G_F n_e E=7.56\times 10^{-5}eV^2(\frac{\rho}{g cm^{-3}})
(\frac{E}{GeV})
\end{equation}
Here, $n_e$ is the electron density and $\rho$ the matter density. 
For anti-neutrinos, we must replace $D$ by $-D$.
For $\rho(g\ cm^{-3})E(GeV)\approx 40$, the effective mass parameter
$D$ is of the order of the mass splitting that is derived from the
atmospheric neutrino anomaly. We then expect matter effects to
be important.

\begin{figure}[tb]
\begin{center}
\epsfig{file=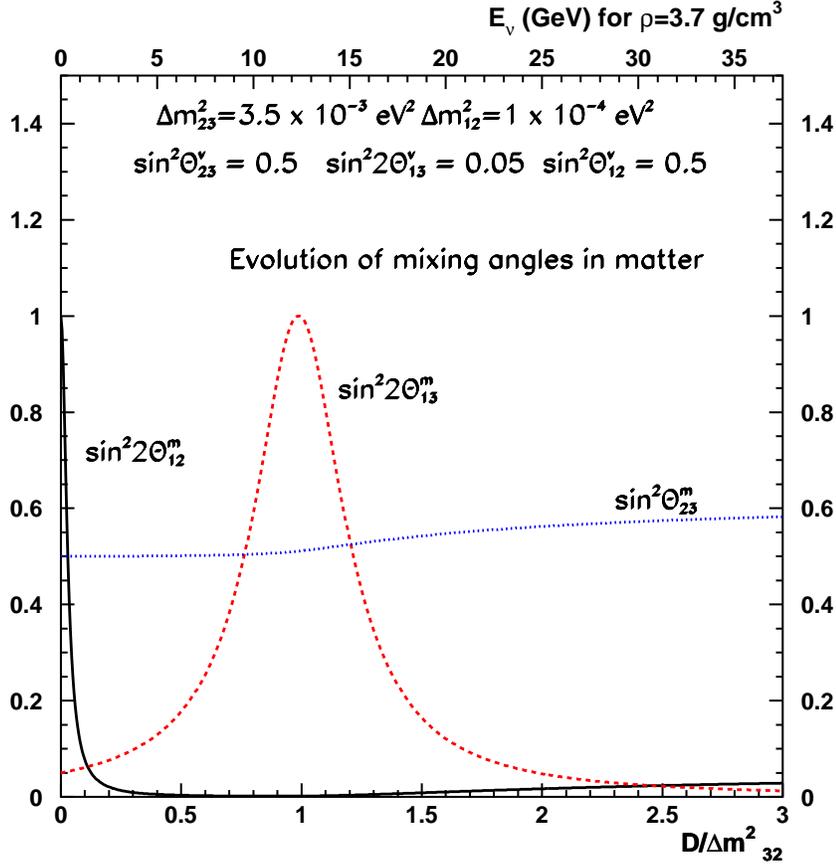,width=12.cm}
\caption{Evolution of mixing angles in matter as a function of 
$D(=2\sqrt{2}G_F N_e E_\nu)/\Delta m^2_{32}$. The reference vacuum
parameters are given in the figure.}
\label{fig:anglemat}
\end{center}
\end{figure}

Within the two-family mixing scheme, the modification of the
flavor transition in matter is taken into account by the mixing angle 
in matter $\theta_m$, which is:
\begin{equation}
\sin^2 2\theta_m(D)=\frac{\sin^2 2\theta}{\sin^2 2\theta+ (
\frac{D}{\Delta m^2}-\cos 2\theta)^2}
\end{equation}
For neutrinos, a resonance condition will be met when 
$D\simeq \Delta m^2\cos2\theta$
and the oscillation amplitude will reach a maximum. 
The resonant neutrino energy $E^{res}$ is
\begin{equation}
\label{eq:resoene}
E^{res}\approx \frac{1.32\times10^4\cos2\theta\Delta m^2(eV^2)}{\rho(g/cm^3)}
\end{equation}

Rather than two-family mixing, we have adopted throughout this 
study three-family framework.
In this context, we use the analytic expressions for the 
matter mixing angles and mass eigenvalues calculated 
in~\cite{zaglauer,pakvasa} (see Appendix A). 

The mass eigenvalues in matter $M_1$, $M_2$ and $M_3$ are: 
\begin{eqnarray}
M_1^2 & = & m_1^2 +\frac{A}{3} - \frac{1}{3}\sqrt{A^2-3B}S -
\frac{\sqrt{3}}{3}\sqrt{A^2-3B}\sqrt{1-S^2}\\
M_2^2 & = & m_1^2 +\frac{A}{3} - \frac{1}{3}\sqrt{A^2-3B}S +
\frac{\sqrt{3}}{3}\sqrt{A^2-3B}\sqrt{1-S^2}\\
M_3^2 & = & m_1^2 +\frac{A}{3} + \frac{2}{3}\sqrt{A^2-3B}S
\end{eqnarray}
where $A$, $B$ and $S$ are given in the Appendix.
For the mixing angles in matter the analytical expressions read: 
\begin{eqnarray}
\sin^2\theta^m_{12} & = & \frac{-(M_2^4-\alpha M_2^2+\beta)\Delta M^2_{31}}
{\Delta M^2_{32}(M_1^4 -\alpha M_1^2+\beta)-\Delta
M^2_{31}(M^4_2-\alpha M_2^2 + \beta)} \\
\sin^2\theta^m_{13} & = & \frac{M_3^4-\alpha M_3^2+\beta}{\Delta
M^2_{31}\Delta M^2_{32}} \\
\sin^2\theta^m_{23} & = &
\frac{G^2s^2_{23}+F^2c^2_{23}+2GFc_{23}s_{23}c_\delta}{G^2+F^2} 
\end{eqnarray}
where $\alpha$, $\beta$, $G$ and $F$ are found in the Appendix.

To illustrate matter effects in three-neutrino mixing framework,
we show in Figure~\ref{fig:anglemat} the values of the mixing angles
in matter, plotted as a function of $D/\Delta m^2_{32}$, or equivalently
of $\rho\times E$. The parameter values in vacuum correspond
to our reference values for atmospheric and LMA-MSW solar experiments.
The resonant behavior of $\theta^m_{13}$ is clearly visible.
It gives maximum oscillation at a neutrino energy
of about 12~GeV for a density of 3.7~$g\ cm^{-3}$. There is a similar
resonant behavior for $\theta^m_{12}$ but it occurs at low energy
since it is driven by $\Delta m^2_{21}$. For $D>\Delta m^2_{32}$,
the angles $\sin^2 2\theta^m_{12}$ and $\sin^2 \theta^m_{23}$
tend to rise slightly, because the non-vanishing $\Delta m^2_{21}$ splitting
removes the degeneracy between muon and tau flavors.

Our results have been computed assuming 
a constant density along the whole neutrino path,
and equal to the mean density, obtained integrating over the earth profile
\cite{eprof}.
This approximation
yields, as shown in e.g. Ref.~\cite{shrock}, similar results to those obtained
by numerical integration using the actual Earth's density profile.

The oscillation probability through matter
will be a function of eight parameters:
(a) the vacuum three mixing angles $\theta_{12}$,$\theta_{13}$,$\theta_{23}$;
(b) the vacuum two mass differences squared $\Delta m^2_{12}$, $\Delta m^2_{23}$;
(c) the average earth density $\rho$;
(d) the baseline L;
(e) the neutrino energy $E$.

\section{Choice of baseline and event rates}
\label{sec:evratedist}

The exact parameters of a neutrino factory are not
yet completely fixed and realistic scenarios are in 
the process to be defined\cite{nufacwww}. 
However, based on \cite{nufacwww}, we can assume that the muons 
in the storage ring have an energy
$E_\mu = 30$ GeV and that after one year of 
operation, the factory should deliver about
$10^{20}$ ``useful'' muons decays of both polarities in the straight section 
pointing towards the far detector location. We base
our ultimate reach on $10^{21}$ ``useful'' muons decays.
Even an integrated intensity of $10^{22}$ might be eventually reachable.

We compute the fluxes assuming unpolarized muons and
disregarding muon beams divergences within the storage
ring. We integrate the expected event rates using
a neutrino-nucleon Monte-Carlo generator~\cite{nux}.
The total charged current (CC) cross
section is technically subdivided into
three parts: the exclusive
quasi-elastic scattering channel $\sigma_{QE}$ and
the inelastic cross section $\sigma_{inelasic}$ which
includes all other processes except charm
production which is included separately.

\begin{table}[htb]
\begin{center}
\begin{tabular}{|cc|c|c|c|c|c|c|}
\hline
\multicolumn{8}{|c|}{Event rates for various baselines} \\ \hline\hline
 & & \multicolumn{2}{|c|}{L=732 km} & \multicolumn{2}{|c|}{L=2900 km} & 
\multicolumn{2}{|c|}{L=7400 km} \\
\cline{3-8}
 & & $N_{tot}$ & $N_{qe}$ & $N_{tot}$ & $N_{qe}$ & $N_{tot}$ & $N_{qe}$ \\\hline
 & $\numu$ CC & 226000 & 9040 & 14400 & 576 & 2270 & 90 \\
$\mu^-$ & $\numu$ NC &  67300 & $-$ &  4120 & $-$ & 680 & $-$  \\
$10^{20}$ decays & $\anue$ CC &  87100 & 3480 & 5530 & 220 & 875 & 35 \\
 & $\anue$ NC &  30200 & $-$  & 1990 & $-$  &  300 & $-$  \\ \hline\hline
 & $\anumu$ CC & 101000 & 4040 & 6380 & 255 & 1000 & 40 \\
$\mu^+$ & $\anumu$ NC &  35300 & $-$ & 2240 & $-$ &  350 & $-$ \\
$10^{20}$ decays & $\nue$ CC &  197000 & 7880 & 12900 & 516 & 1980 & 80 \\
 & $\nue$ NC &  57900 & $-$ & 3670 & $-$ &  580 & $-$ \\ \hline
\end{tabular}
\caption{Expected events rates for a 10 kton (fiducial) detector in
case no oscillations occur for $10^{20}$ muon decays. $N_{tot}$ is
the total number of events and $N_{qe}$ is the number
of quasi-elastic events.}
\label{tab:rates}
\end{center}
\end{table}

Table~\ref{tab:rates} summarizes the expected rates for
the 10 kton fiducial mass and $10^{20}$ muon decays (expected 1 year
of operation). 
$N_{tot}$ is
the total number of events and $N_{qe}$ is the number
of quasi-elastic events.

Even though our study is site non-specific, the chosen baselines
could correspond to the distances between
the Laboratori
Nazionale del Gran Sasso (LNGS) and neutrino factories
at (1) CERN ($L=732$ km, $<\rho_{Earth}> = 2.8$ g/cm$^3$), 
(2) Canary Islands ($L=2900$ km, $<\rho_{Earth}> = 3.2$ g/cm$^3$) and 
(3) Fermilab ($L=7400$ km, $<\rho_{Earth}> = 3.7$ g/cm$^3$).

\section{The four main classes of events}
\label{sec:evtclass}

Muon identification, charge and momentum
measurement provide discrimination between $\nu_\mu$ 
and $\bar\nu_\mu$ charged current (CC) events. Good
$\nu_e$ CC versus $\nu$ NC
discrimination relies on the fine granularity of the target. Finally, the 
identification of $\nu_\tau$ CC events requires a precise measurement 
of all final state particles.

It is natural to classify the events in four classes\cite{us1}. We
illustrate them for the case of $\mu^-$ stored in the ring. 
\begin{enumerate}
\item {\bf Right sign muons ($rs\mu$):} the leading muon has the
same charge as those circulating inside the ring. Their origin
is from
\begin{enumerate}
\item non-oscillated $\numu$ CC
\item $\numu\ra\nutau$ CC, $\tau^-\ra\mu^-$ decays
\item hadron decays in neutral currents.
\end{enumerate}
\item {\bf Wrong sign muons ($ws\mu$):} the leading muon has
opposite charge to those circulating inside the ring.
Opposite-sign leading muons can only be produced by neutrino oscillations, since
there is no component in the beam that could account for them.
\begin{enumerate}
\item $\bar{\nu}_e\to\bar{\nu}_\mu$ oscillations
\item $\bar{\nu}_e\to\bar{\nu}_\tau$ oscillations, $\tau^+\ra\mu^+$ decays
\item hadron decays in neutral currents.
\end{enumerate}
\item {\bf Electrons ($e$):} events with a prompt electron and no primary
muon identified.
Events with leading electron or positron are produced by the 
charged-current interactions of the
following neutrinos:
\begin{enumerate}
\item non-oscillated $\bar{\nue}$ neutrinos
\item $\numu\ra\nue$ oscillations
\item $\bar{\nu}_e\to\bar{\nu}_\tau$ or $\numu\ra\nutau$ oscillations
with $\tau\ra e$ decays
\end{enumerate}
\item {\bf No Lepton ($0\ell$):} events corresponding to NC interactions or
$\nu_\tau$ CC events followed by a hadronic decay of the tau lepton.
Events with no leading electrons or muons will be used to 
study the $\numu\ra\nutau$ oscillations. These events 
can be produced in 
\begin{enumerate}
\item neutral current processes
\item $\bar{\nu}_e\to\bar{\nu}_\tau$ or $\numu\ra\nutau$ oscillations
with $\tau\ra hadrons$ decays
\end{enumerate}
\end{enumerate}

The last two classes can only be cleanly studied in
a fine granularity detector.

The most effective way to fit the oscillation parameters is
to study the visible energy distribution of
the four classes of events defined above, since assuming
the unoscillated spectra are known, they contain
direct information on the oscillation probabilities.

Of course,
for electron or muon charged current events, the visible
energy reconstructs the incoming neutrino energy. In the
case of neutral currents or the charged current of tau
neutrinos, the visible energy is less than the visible
energy because of undetected neutrinos in the final state.
The information is in this case degraded but can
still be used.

Our analyses are performed on samples of fully generated Monte-Carlo\cite{nux} 
events, which include proper kinematics of the events, full hadronization
of the recoiling jet and proper exclusive polarized tau decays when 
relevant\footnote{Our simulation has been bench-marked on the comparison
of the kinematic features of the lepton and hadronic jet of real
neutrino data accumulated in the NOMAD experiment 
(see e.g. \cite{icanoe}).}.
Nuclear effects, which are taken into
account by the FLUKA model~\cite{fluka}, are included
as they are important for a proper estimation of the tau
kinematical identification.

The detector response is included in our analyses using a fast simulation which
parameterizes the momentum and angular resolution of the emerging 
particles, using essentially the following values:
electromagnetic shower $3\%/\sqrt{E}\oplus 1\%$, 
hadronic shower $\approx 20\%/\sqrt{E}\oplus 5\%$, and magnetic muon
momentum measurement $20\%$.

\begin{figure}[tb]
\begin{center}
\begin{minipage}{7.5cm}
\epsfig{file=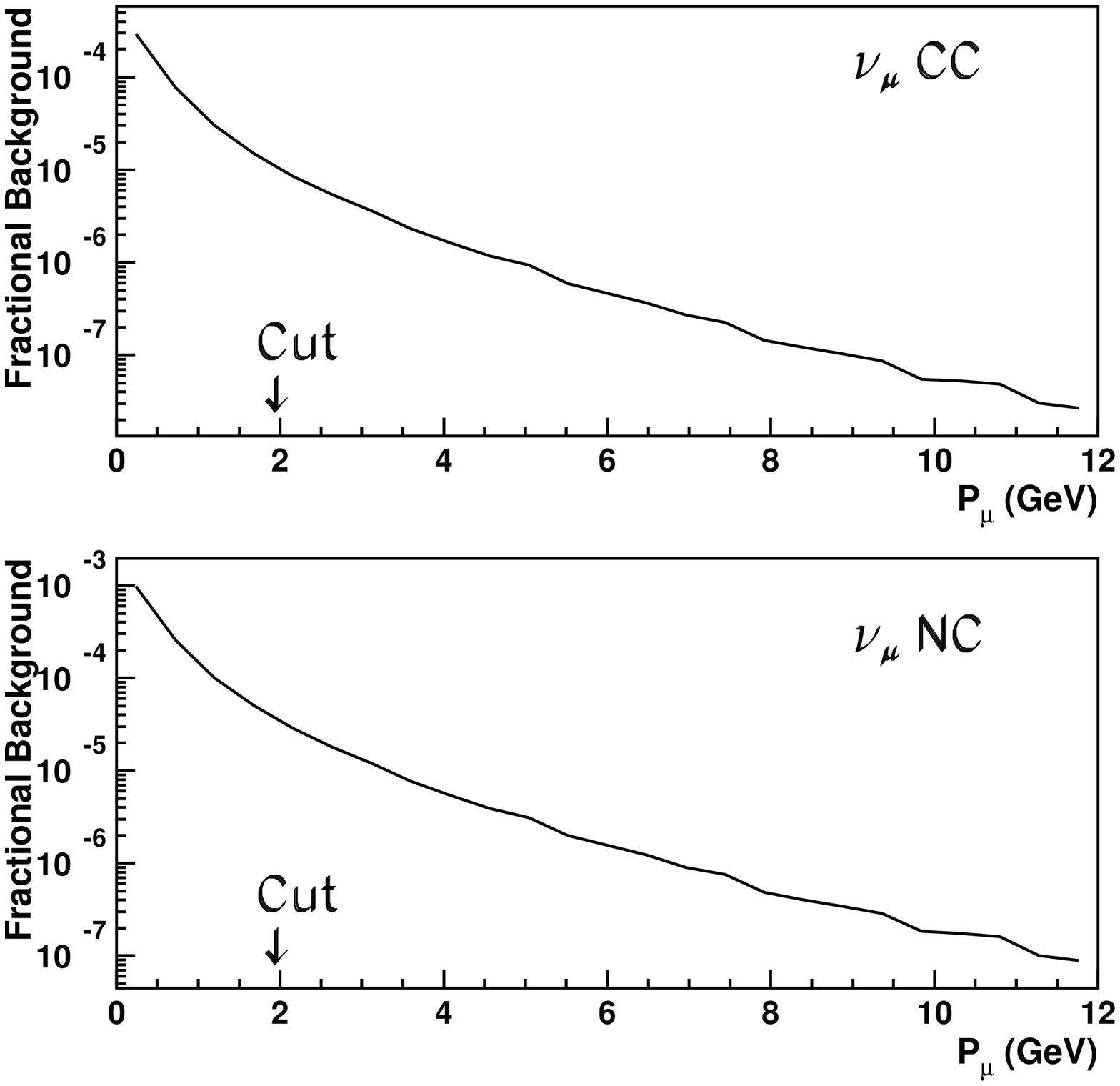,width=8.cm}
\end{minipage}
\begin{minipage}{7.5cm}
\epsfig{file=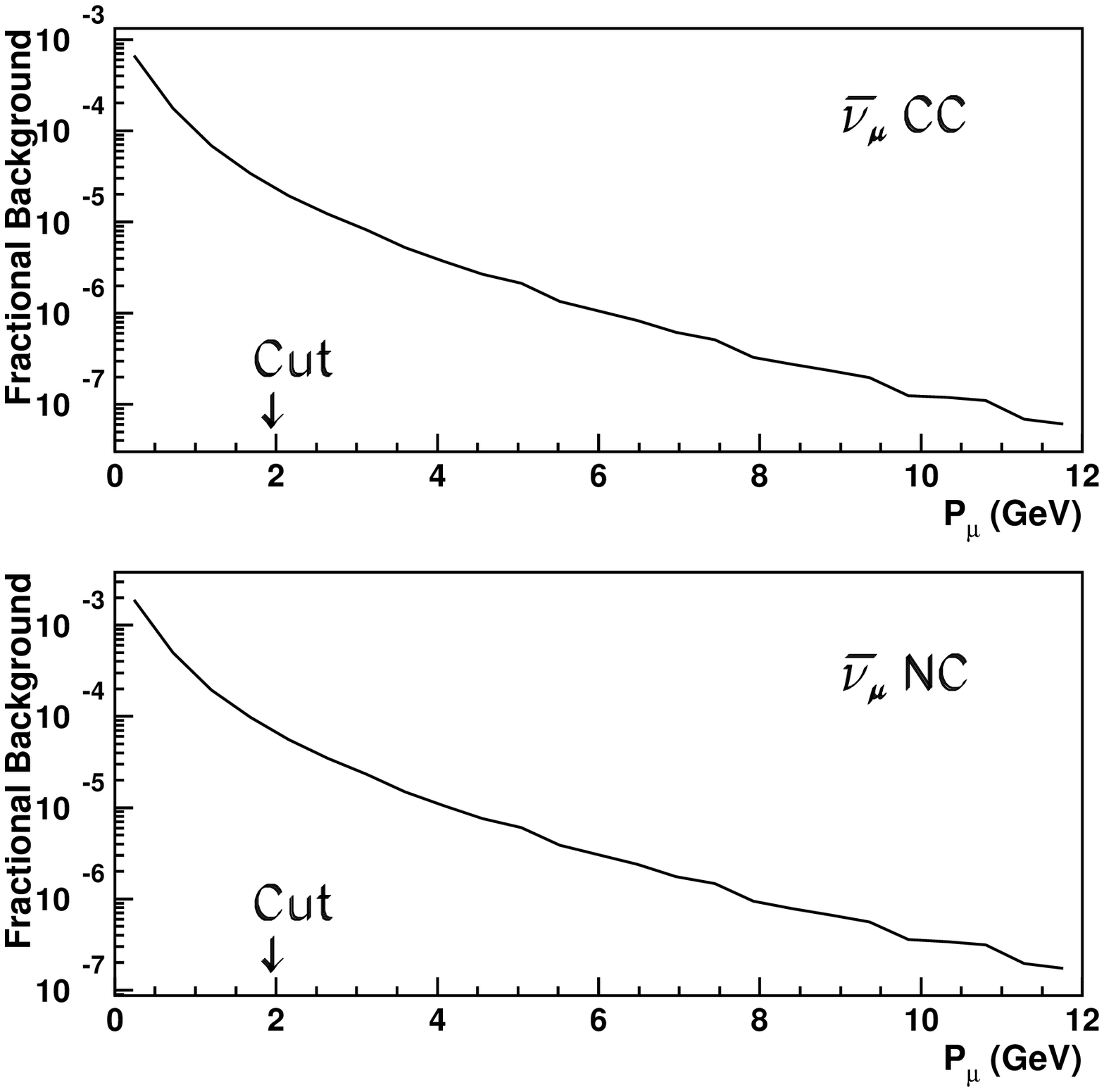,width=8.cm}
\end{minipage}
\caption{Fractional background due to the decay of charged mesons as a 
function of the measured muon momentum.}
\label{fig:bckgnd}
\end{center}
\end{figure}

Hadron decay background can be quite large in a low
density target and could be quite dangerous. Fortunately,
it can be easily suppressed 
by a cut on the muon candidate momentum,
$P_\mu > 2$ GeV, which reduces it to a tolerable level. 
Figure~\ref{fig:bckgnd} illustrates the relative background expected for 
$\nu_\mu$ and $\bar{\nu}_\mu$ NC and CC processes as a function of the muon
momentum. After the cut, the expected contamination for $\nu_\mu$ CC
events is at the level of $10^{-5}$. Real charged current
events maintain an efficiency above $95\%$.

\begin{figure}
\begin{minipage}{7.5cm}
\epsfig{file=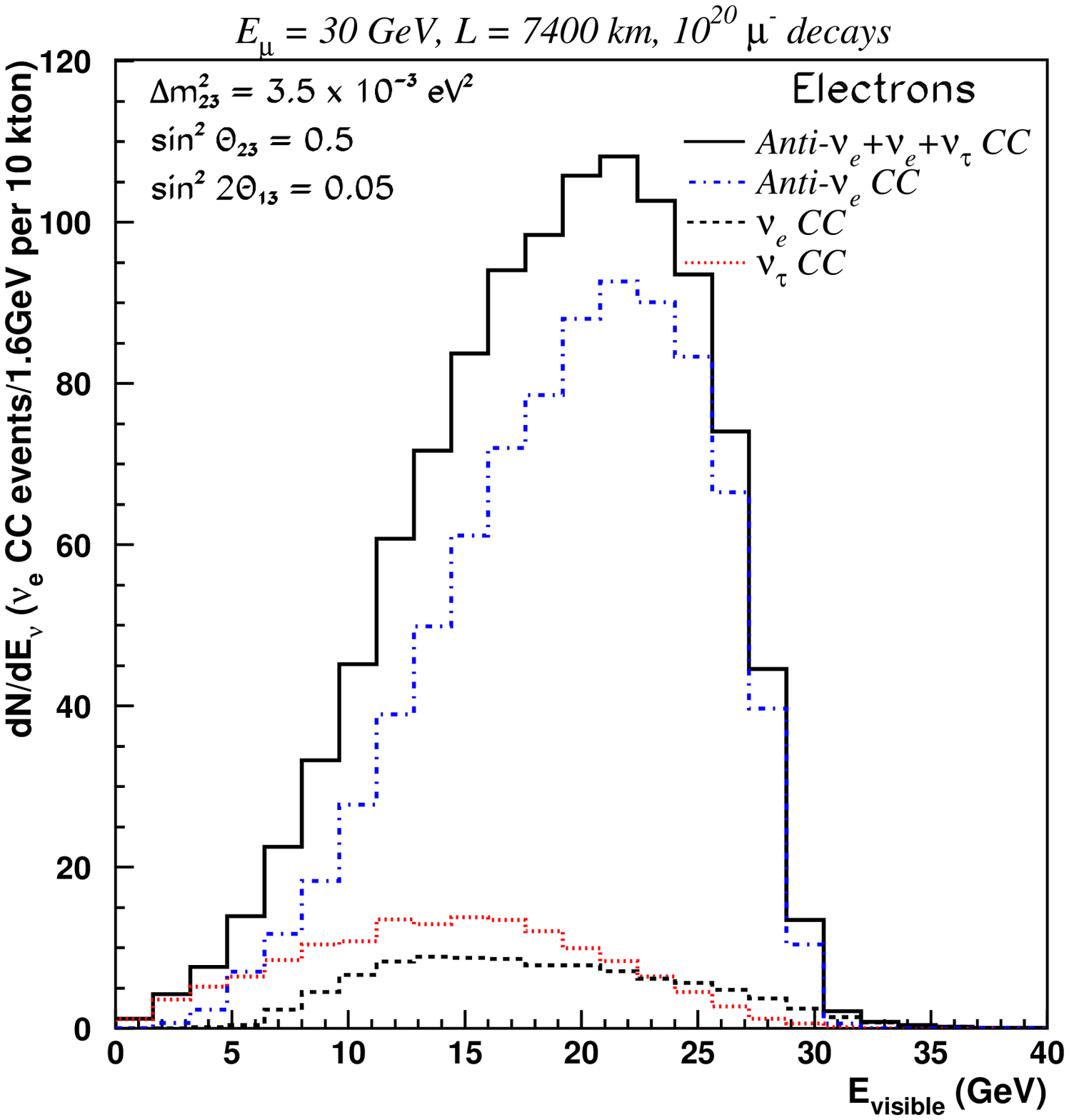,width=7.cm}
\caption{Visible energy spectrum for electron events:
$\nu_e$ CC (dashed line),  $\nu_\tau$ and
$\bar\nu_\tau$ (dotted line) and $\bar{\nu}_e$ CC (dot-dashed). 
The solid
histogram shows the sum of all contributions.}
\label{fig:elecont}
\end{minipage}
\begin{minipage}{7.5cm}
\epsfig{file=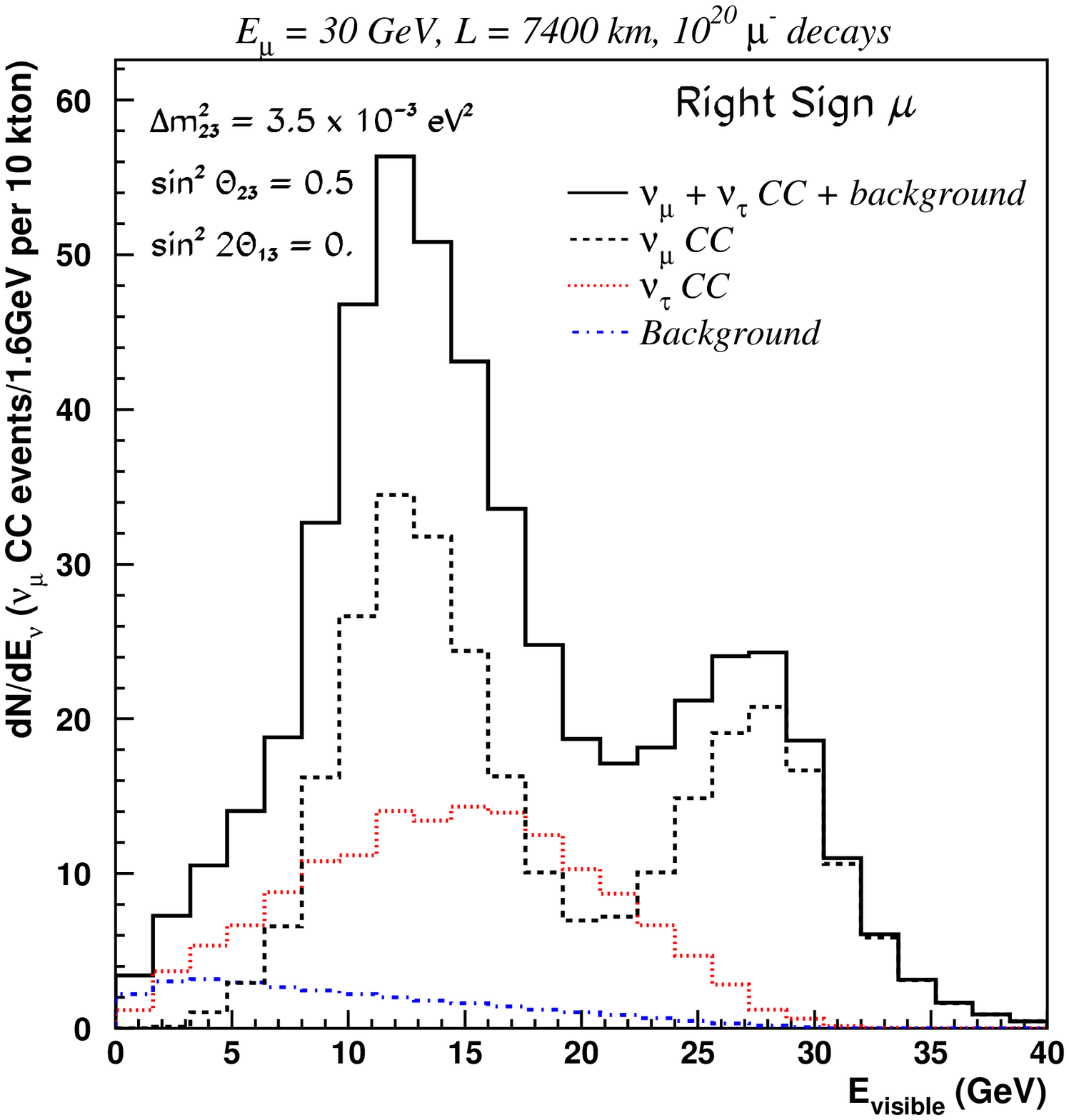,width=7.cm}
\caption{
same as Figure~\ref{fig:elecont} for right-sign muon
sample: $\nu_\mu$ CC (dashed line),  $\nu_\tau$ and
$\bar\nu_\tau$ (dotted line) and meson decay background (dot-dashed). 
The solid
histogram shows the sum of all contributions.}
\label{fig:dipevol}
\end{minipage}
\begin{minipage}{7.5cm}
\epsfig{file=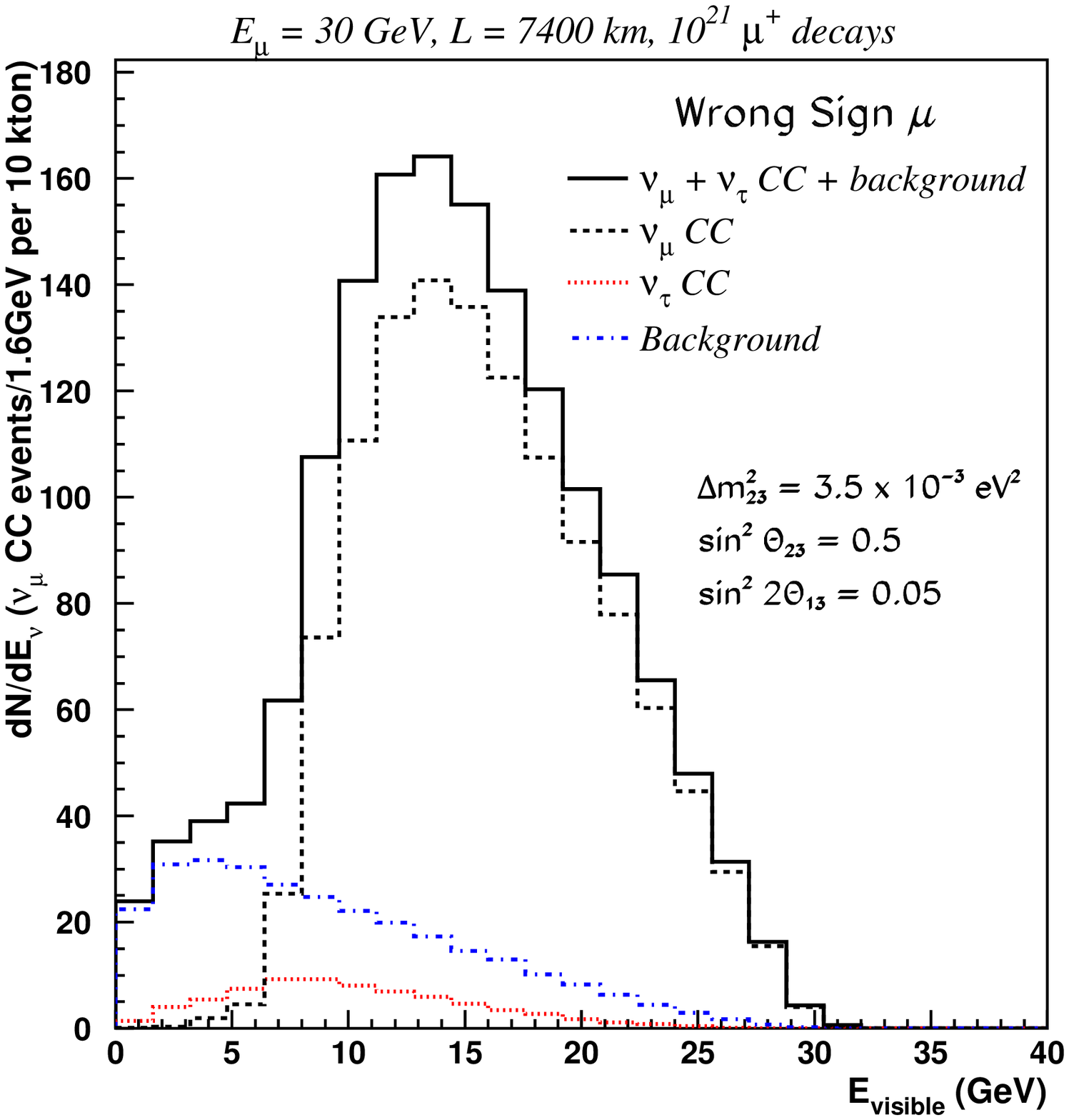,width=7.cm}
\caption{
same as Figure~\ref{fig:dipevol} for wrong sign muon
sample: $\nu_\mu$ CC (dashed line),  $\nu_\tau$ and
$\bar\nu_\tau$ (dotted line) and meson decay background (dot-dashed). 
The solid
histogram shows the sum of all contributions.}
\label{fig:wrongmu}
\end{minipage}
\begin{minipage}{7.5cm}
\epsfig{file=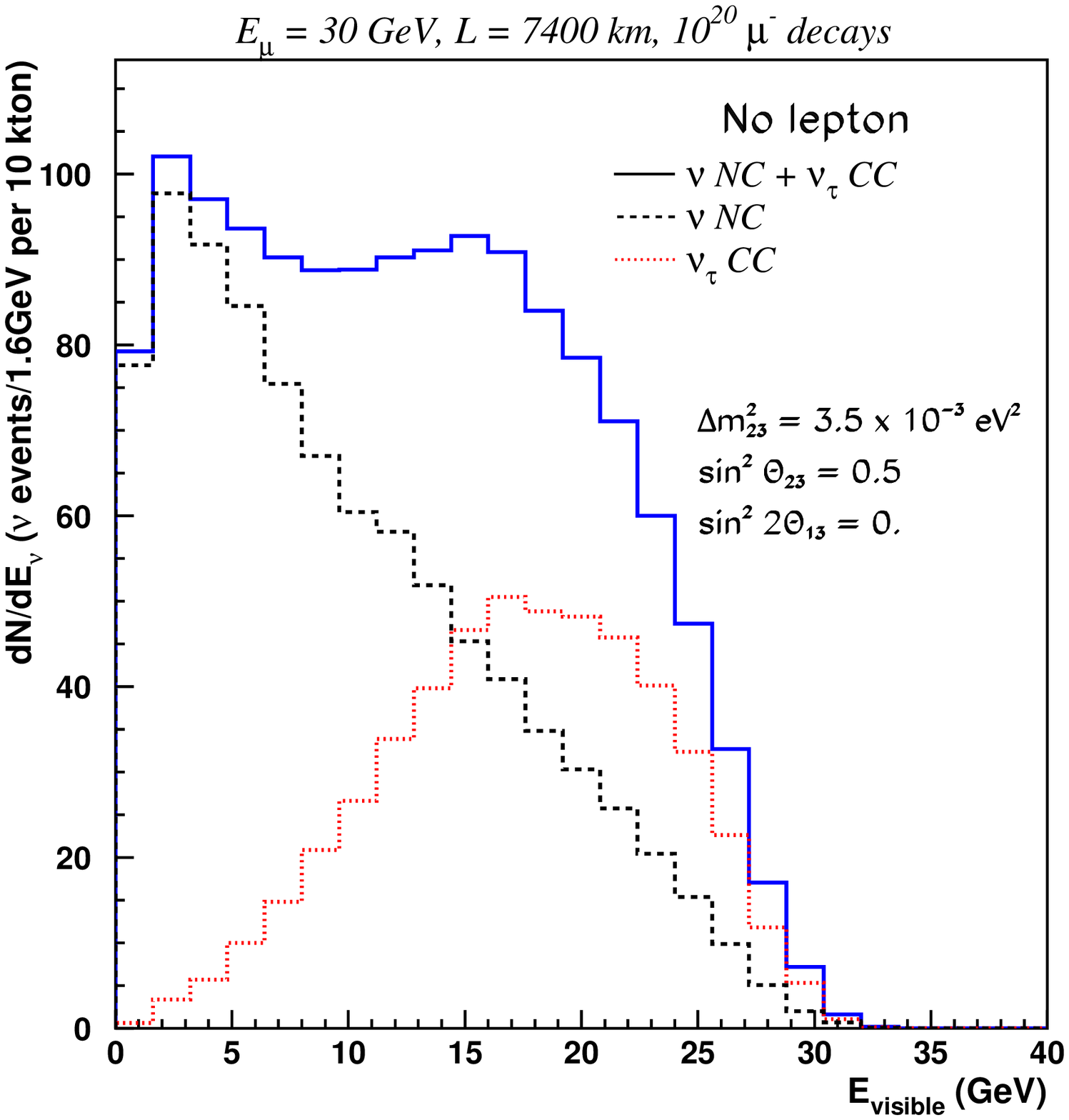,width=7.cm}
\caption{
same as Figure~\ref{fig:dipevol} for the no-lepton
sample: $\nu$ NC (dashed line),  $\nu_\tau$ and
$\bar\nu_\tau$ (dotted line).
The solid
histogram shows the sum of all contributions.}
\label{fig:nccont}
\end{minipage}
\end{figure}

The visible energy is computed as the modulus of the vector sum of the
momenta of each visible particle in the event.
Figures~\ref{fig:elecont},~\ref{fig:dipevol},~\ref{fig:wrongmu} 
and~\ref{fig:nccont} show the reconstructed visible energy
at the baseline $L=7400$km normalized to $10^{20}\mu$'s
for each event class for a specific oscillation scenario
with $\Delta m_{32}^2 = 3.5 \times 10^{-3}$ eV$^2$, 
$\sin^2 \tet23 = 0.5$ and $\sin^2 2\theta_{13} = 0.05$.
The different contributions including backgrounds for each 
event class have been evidenced in the plots. 
For example, in Figure~\ref{fig:dipevol}, the different processes that
contribute to the right-sign muon class are
unoscillated muons, taus and background events. 

\section{Further classification based on kinematical analysis}
\label{sec:kineanal}

\begin{table}
\begin{center}
\begin{tabular}{|c|c|c||c|c|c|}
\hline
\multicolumn{6}{|c|}{$\nu_\mu\to\nu_\tau$ appearance search} 
\\ \hline\hline
Cuts & $\tau\to l$ & CC background & Cuts & $\tau\to h$  & NC background \\ \hline
Initial & $100\%$ & $100\%$ & Initial & $100\%$ & $100\%$ \\\hline\hline
\multicolumn{6}{|c|}{Loose cuts} \\ \hline 
$P_T^{l}<0.5$GeV & $50\%$ & $14\%$ & $P_T^{miss} < 1$GeV & $72\%$ & $40\%$ \\
$P_T^{miss} > 0.6$GeV & $40\%$ & $0.5\%$ & $Q_T > 0.5$GeV & $30\%$ & $2\%$ \\\hline\hline
\multicolumn{6}{|c|}{Tight cuts} \\ \hline 
$P_T^{l}<0.5$GeV & $50\%$ & $14\%$ & $P_T^{miss} < 1$GeV & $72\%$ & $40\%$ \\
$P_T^{miss} > 1$GeV & $20\%$ & $0.08\%$ & $Q_T > 1$GeV & $6\%$ & $0.07\%$ \\\hline
\end{tabular}
\caption{$\nu_\tau$ appearance search for the leptonic and hadronic decay 
modes of the tau lepton. Overall signal efficiencies and fractional 
remaining backgrounds are quoted. The case labelled as ``Tight cuts'' 
correspond to the situation where one event background is expected, for 
$10^{20}$ muon decays, at the farthest location ($L=7400$ km).}
\label{tab:tausearch}
\end{center}
\end{table}

An efficient identification of $\nu_\tau$ induced charged current 
events requires a precise measurement of all final state particles. 
Excellent calorimetry allows to take full 
advantage of the special kinematic features of $\nu_\tau$
events. We independently search for the leptonic and hadronic tau
decay modes. 

\begin{figure}
\begin{center}
\epsfig{file=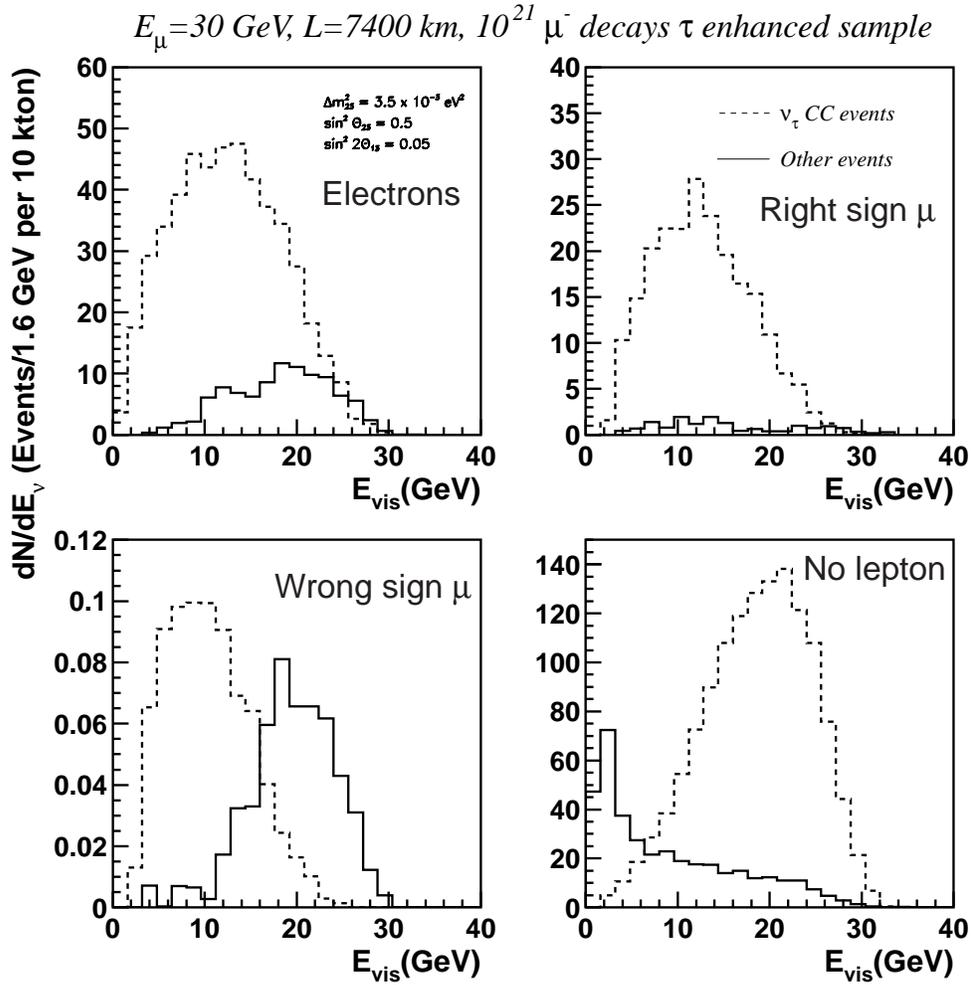,width=14.cm}
\caption{Visible energy spectrum for the four event classes after application of
loose kinematic cuts (see text). The presence of 
$\tau$ events in the third class (wrong-sign muons) is an indication of the 
process $\nu_e\to\nu_\tau$.}
\label{fig:tauenhance}
\end{center}
\end{figure}

For the $\tau\to l\nu\nu$ decay mode, the main background comes from 
$\nu_l$ CC. To enhance the separation between $\tau$ and 
background events, we demand the event
missing $P_T$ to be larger than 0.6 GeV and the transverse momentum of 
the lepton candidate,$P_T^{l}$, to be smaller than 0.5 GeV. 
This set of cuts is referred to as ``loose cuts'' and it will be used 
to perform a check on 
appearance/disappearance consistency (see subsection~\ref{sec:taus}). 
In Table~\ref{tab:tausearch}, we show that the overall $\tau$ efficiency 
for ``loose cuts'' is $40\%$ for a CC background level of 
$\sim 5\times10^{-3}$. Figure~\ref{fig:tauenhance} shows the energy spectra
for the four event classes after application of these cuts, for $\nu_\tau$
CC and other types of events. No energy cut has been applied, but the fact of
using the energy spectra in the fit also exploits the difference in energy
spectra of $\nu_\tau$ CC events.\par

A set of ``tight cuts'' is also applied, aiming at 
having one expected background event at the farthest location in case 
$10^{20}$  ``useful'' muons decays are delivered. As we can see from 
Table~\ref{tab:tausearch}, the overall tau efficiency in this case amounts 
up to $20\%$.

For hadronic
decays the most important source of background correspond to NC events. 
If we demand a $P_T^{miss}$ smaller than 1 GeV
and a transverse momentum of the hadron candidate with respect to the
total event momentum, $Q_T$, larger than 0.5 GeV (``loose cuts'') 
only $2\%$ of the initial background survives for an overall tau 
efficiency of $30\%$. If we require only one NC background event survivor 
at $L=7400$ km (``tight cuts''), the signal efficiency drops to $6\%$ 
due to the stringent $Q_T$ requirement imposed. 

\section{Quasi-elastic final states}
\label{sec:quasi}
The quasi-elastic process, while rare, is a clean process
that allows to separate neutrino from anti-neutrino events,
in principle for all neutrino flavors, since
$\nu_\ell + n \ra \ell^- + p$
and $\bar\nu_\ell + p \ra \ell^+ + n$.
The recoil proton is easily identifiable within the
high-granularity target. 

This channel is particularly interesting to study oscillation
in the electron channel. Starting from negative muons
circulating in the storage ring, we look for exclusive electron-proton
final states. These provide ``background-free'' oscillation signals:
\begin{eqnarray}
\mu^-\ra e^-& \bar\nue&\numu \nonumber \\ 
& & \ra \nue + n \ra e^- + p\\
& & \ra \nutau + n \ra\tau^- + p \ra e^-\nu\nu + p
\end{eqnarray}
since $\bar\nue +p \ra e^+ + n$.

The selection of quasi-elastic events is the only way
to identify the helicity of neutrino electrons in absence
of measurement of the electron charge.

\section{Oscillation parameters fitting}
\label{sec:fit}

Given the adopted parameterization of the mixing matrix,
we have a priori a total of 7 free parameters, which can be
represented by the vector:
\begin{equation}
\vec P = (\Delta m^2_{21},\Delta m^2_{32},\sin^2 \theta_{12},
\sin^2 2\theta_{13}, \sin^2 \theta_{23}, \delta, \rho)
\end{equation}

The values of the parameters governing the oscillations are extracted
from a global fit of the visible energy distributions obtained for each event 
class. The fit is performed with the MINUIT \cite{minuit} package, and is 
expected to get back the same values of the parameters, starting from the reference
distributions. 

At each iteration, a different 
set of parameters is probed, and with the same procedure used to get 
the reference histograms.

For a given polarity $\lambda$ of the muons in the storage ring, we compute
$\chi^2$'s of the 
difference between the binned oscillated spectra, which will be function
of the parameters, and the reference histograms. We define a $\chi^2$ for
each of the four classes of events, i.e. the electrons ($e$), the right-sign muon ($rs\mu$),
the wrong sign muons ($ws\mu$) and the no lepton class ($0\ell$):
\begin{equation}
\chi^2_{\lambda, all}= \chi^2_{\lambda,e}+\chi^2_{\lambda,rs\mu}+\chi^2_{\lambda,ws\mu}+\chi^2_{\lambda,0\ell}
\end{equation}
where
\begin{equation}
\chi^2_{\lambda,c}(\vec P)=\sum_{i} \left(\frac{N^c_{i}(\vec
P)-N_{i}^{c}(\vec P_{ref})}{\sigma^2_{i,c}} \right)^2.
\end{equation}
The sum runs over 25 equally spaced energy bins ($E_i$, $i=1,25$).
Here, $N^{c}_i(\vec P_{ref})$ is the expectation for the reference values 
and $N^c_i(\vec P)$ is the ``data'' obtained for a given set of the
oscillation parameters $\vec P$. $\sigma_{i,c}$ contains both
statistical and systematic contributions to the total error.
In order to improve the  statistical treatment of bins with low statistics, 
a bin that possesses less than 40 events is assumed to be
Poisson distributed and therefore its contribution is
computed as $2(N^c_i(\vec P)-N^{c}_i(\vec P_{ref}))+2 N_i 
\ln(N^c_i(\vec P)/N^{c}_i(\vec P_{ref}))$ (see Ref.~\cite{cousins}).

The systematic error takes into account the uncertainties in the
knowledge of the beam, neutrino cross sections and selection
efficiencies and we assume it amounts up to $2\%$ uncorrelated from 
bin to bin. 

We assume that a neutrino factory will operate with alternate
runs of opposite muon polarities, therefore eight energy
distributions can be fitted simultaneously:
\begin{equation}
\chi^2_{all}= \chi^2_{+,all}+\chi^2_{-,all}
\end{equation}

The values of the fitted parameters $\vec P$ are obtained minimizing the 
$\chi^2(\vec P)$.

It is in practice not always possible to fit all the free parameters,
since for some parameter-space regions, the oscillation effects at
the chosen baselines and energies can be negligible. 
In particular, this can be the case for $\Delta m^2_{21}$ and $\sin^2
\theta_{12}$ which drive the solar oscillations. For values
$\Delta m^2_{21}\ll 10^{-4}\ \rm eV^2$, we are insensitive to
the ``solar'' sector.

We therefore adopted successive fitting procedures with an increasing
number of free parameters.

At first, we can simplify the three-family oscillation picture
if the oscillations produced by $\Delta m^2_{21}$ can be neglected
at the considered baselines and energies. 
In this case, only the mass difference squared
$\Delta m^2_{32}$ and the two mixing angles $\theta_{13}$ and $\theta_{23}$ 
are relevant. The oscillation probabilities are given in the Appendix.
For example, for $\nue\ra\numu$ oscillations, it is:
\begin{equation}
P(\nue\ra\numu,E,L)  = 
\sin^2(2\theta^m_{13})\sin^2(\theta^m_{23})\Delta^2_{32}
\end{equation}
where  $\Delta^2_{32} =  \sin^2\left((M_3^2-M_2^2) L/4E\right)$.
The fit has in this case 4 free parameters, which can be
represented by the vector:
\begin{equation}
\vec P_{1ms} = (\Delta m^2_{32},\sin^2 2\theta_{13}, \sin^2 \theta_{23},\rho)
\end{equation}
The $\delta$ phase has disappeared, since within this approximation
it becomes unphysical. The results are presented in
Section~\ref{sec:mixing}, assuming a $\Delta m^2_{32}$ parameter
varying in the range of values favoured by current atmospheric
data ($\Delta m^2_{32} =  3.5,5, 7 \times 10^{-3} \ {\rm eV}^2$), 
a maximal (2-3)-mixing $\sin^2 \tet23 = 0.5$ and a $\tetonethree$ value compatible
with CHOOZ results~\cite{chooz} and recent fits to data~\cite{lisi} 
($\sin^2 2\tetonethree  =  0.05$).

We return to the general three-family scenario in Section~\ref{sec:cp}
where we consider sensitivity to the $CP$ violation.

In order to compute the precision of the determination of the parameters,
we consider two methods: (1) a one-dimensional ``scan'' of a given
parameter; the other variables are left free and mininized at each step;
the resp. 1,2,3 sigmas are given by resp. $\chi^2_{min}+1$, $+4$ and $+9$.
(2) a two-dimensional ``scan'' of a two-parameter plane; the other
variables are left free and minimized at each point in the plane;
the resp. 68\%, 90\%, 99\% C.L. are given by resp.
$\chi^2_{min}+2.3$, $+6.0$ and $+9.2$.

\section{Results for one mass scale approximation}
\label{sec:mixing}

\subsection{Case of two-family mixing : $\theta_{13}=0$}
\label{sec:t23}


\begin{table}[tb]
\begin{center}
\begin{tabular}{|c|c|c|c|c|}
\hline
\multicolumn{5}{|c|}{Two family mixing}\\
\multicolumn{5}{|c|}{$\delta(\sin^2\theta_{23})$ for $\theta_{23} =45^o$,
$\theta_{13}=0$} \\ 
\hline \hline
 & \multicolumn{2}{|c|}{Only right-sign muons} & \multicolumn{2}{c|}{All
classes}\\ 
 & \multicolumn{2}{|c|}{$\chi^2_{\pm,rs\mu}$} 
 & \multicolumn{2}{|c|}{$\chi^2_{\pm,all}$} \\
\hline
$\Delta m^2_{32}$ (eV$^2$) & L=7400 km & L=2900 km 
& L=7400 km & L=2900 km \\ \hline
$7\times 10^{-3}$ & $22\%$   & $8\%$  & $20\%$ & $8\%$\\
$5\times 10^{-3}$ & $12\%$   & $12\%$ & $11\%$ & $10\%$ \\
$3.5\times 10^{-3}$ & $10\%$ & $18\%$ & $10\%$ & $16\%$\\ \hline
\end{tabular}
\caption{Precision in the measurement of the mixing angle assuming two 
family mixing for three possible mass differences and two very large
baselines.In all the cases the precision obtained in the measurement
of $\Delta m_{32}^2$ is $1\%$.}
\label{tab:sin}
\end{center}
\end{table}

The detection of the {\it dip} in the energy distribution of
the right-sign muon sample dominates the
precision on the measurement of the mixing angle $\theta_{23}$ and the mass
difference $\Delta m^2_{32}$. This is true provided that the beam
energy and baseline are chosen in such a way that the $\numu$
disappearance maximum is visible in the oscillated spectrum.
Table~\ref{tab:sin} summarizes the expected accuracies.
With $10^{20}$ muon decays 
of each polarity, precisions of $1-2 \%$ are expected in the
determination of $\Delta m^2_{32}$ while for the 
precision on $\sin^2\theta_{23}$ is around 10$\%$, in agreement with 
results quoted in~\cite{barger}. 

\begin{figure}[tb]
\begin{center}
\epsfig{file=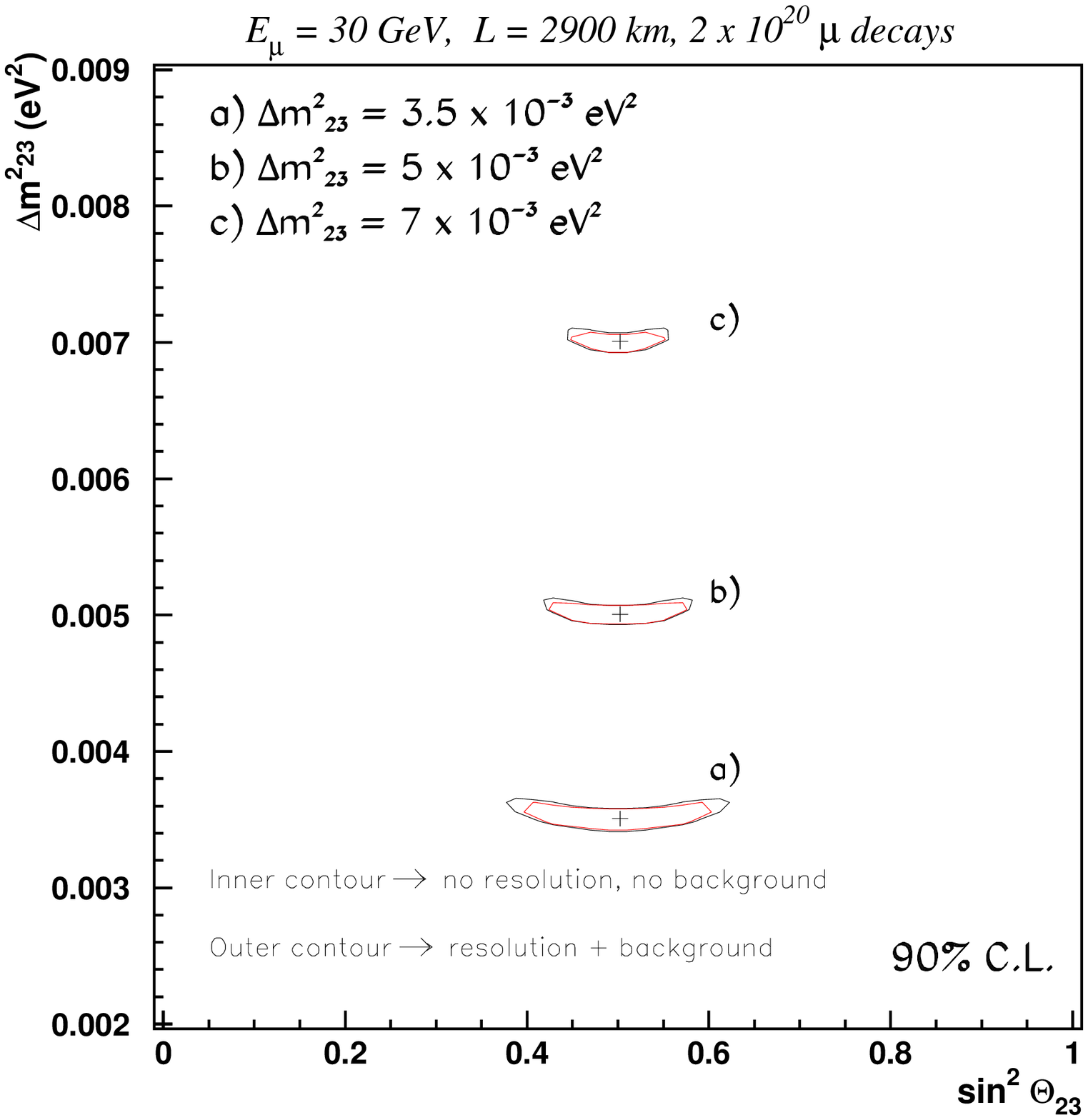,width=6.5cm}
\epsfig{file=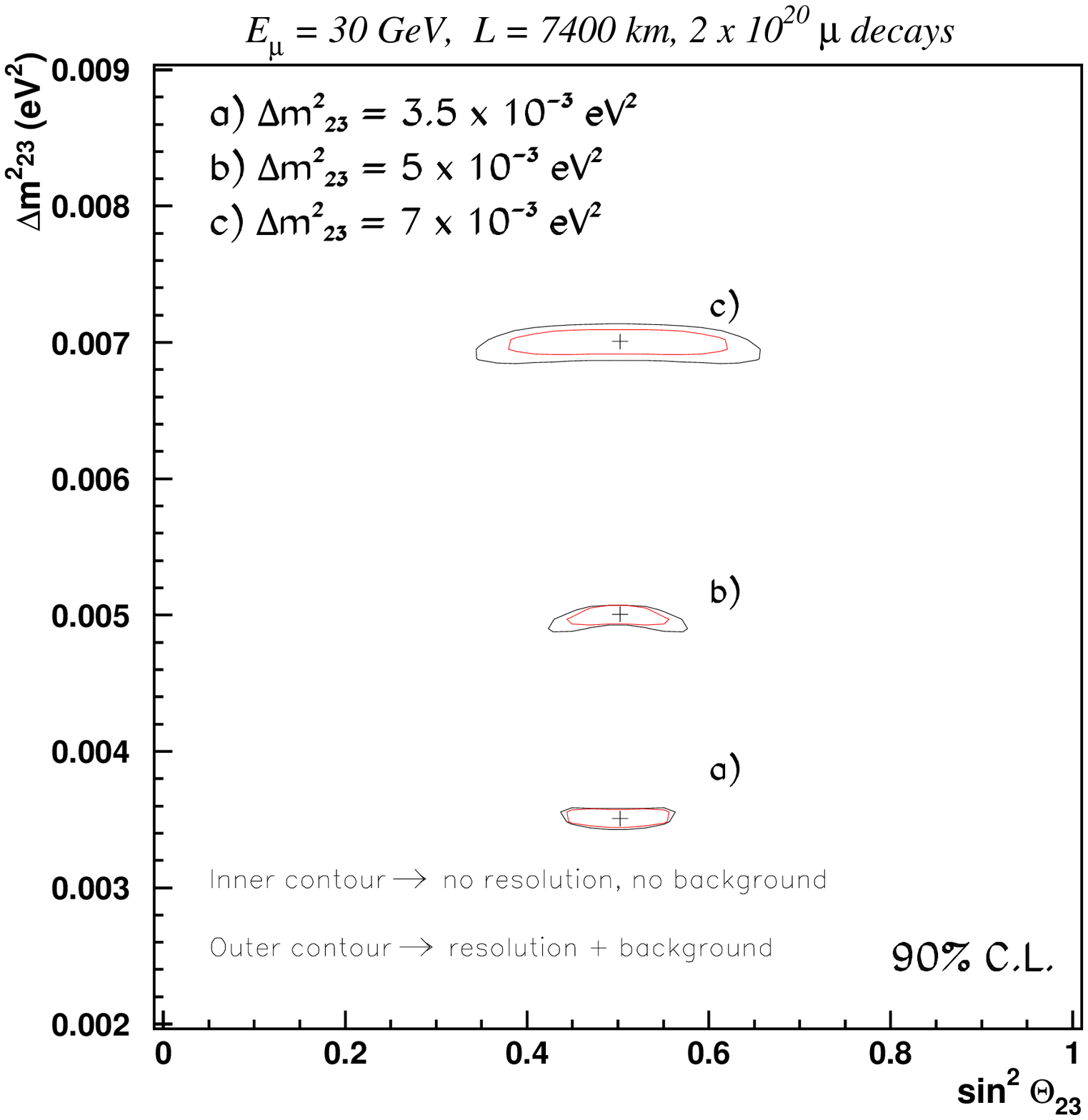,width=6.5cm}
\caption{Two-dimensional 90\%C.L. contours for $\Delta m^2_{32}$ and $\sin^2 \theta_{23}$
for three different $\Delta m^2_{32}$ values. Inner contours: no background
and perfect muon resolution; outer contours: backgrounds and muon
resolution included.}
\label{fig:twofamilynoback}
\end{center}
\end{figure}

We compare in figure~\ref{fig:twofamilynoback} how the precision
on the mass difference and mixing angle changes when experimental 
resolutions and backgrounds are
disregarded. We see that our fits are barely affected 
and only for the cases where the {\it dip} is not seen, 
the instrumental effects and backgrounds spoil at the level of a few
per cent the accuracy on the oscillation parameters.

\subsection{Case of three-family mixing : $\theta_{13}\ne 0$}
\label{sec:t13}

\begin{figure}
\begin{center}
\epsfig{file=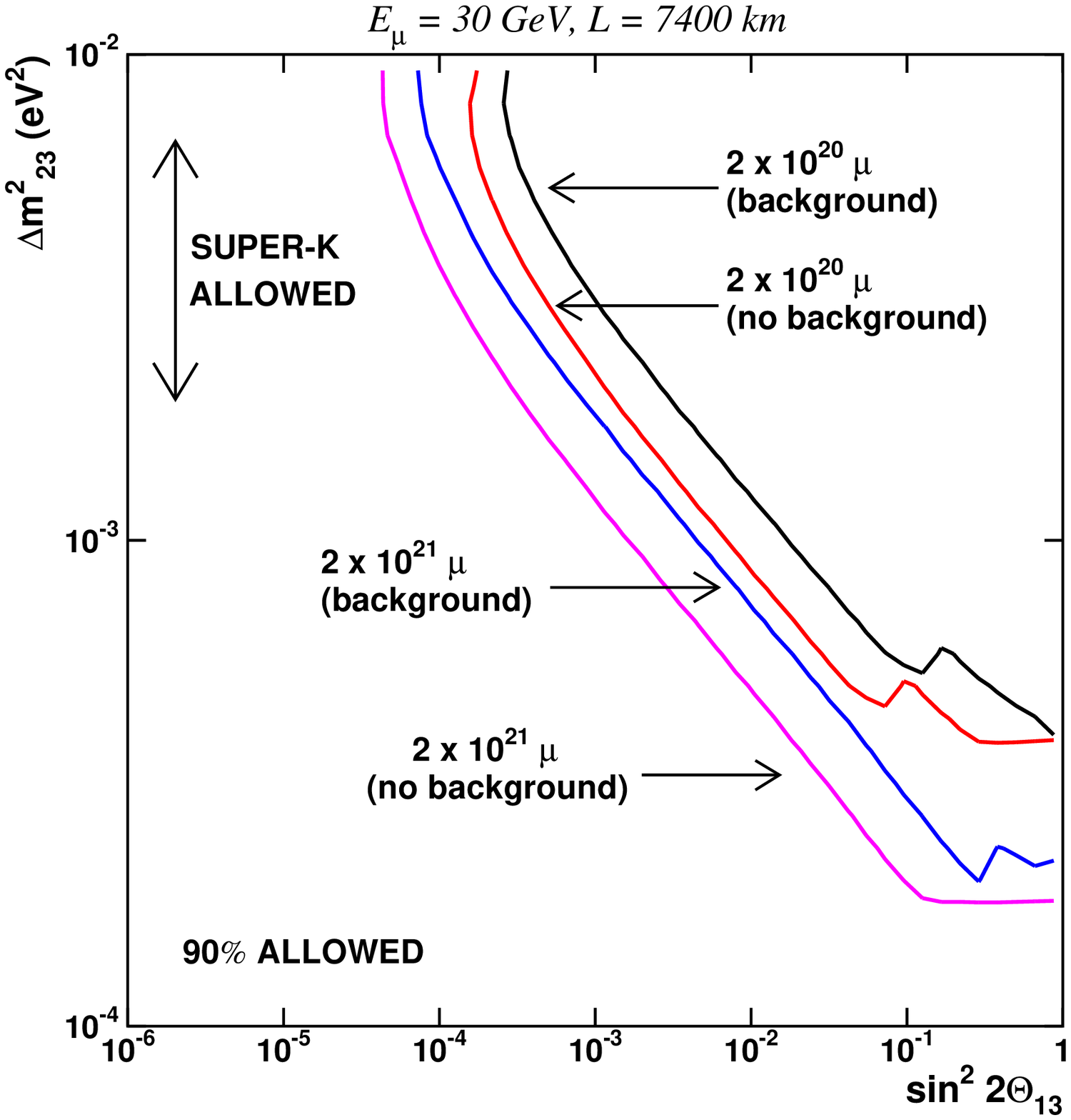,width=14.cm}
\caption{Sensitivity on $\tetonethree$}
\label{fig:sensi}
\end{center}
\end{figure}

In order to obtain the best sensitivity on the mixing angle
$\theta_{13}$, the search for wrong sign muons is ideal, since
it will be a direct signature for $\nue\ra\numu$ oscillations.

For this kind of study, since we are dealing with the smallest number
of signal events, the sensitivity does strongly depend on the
ability of rejecting background. 
Figure~\ref{fig:sensi} shows for $L=7400$ km, 
the sensitivity on $\tetonethree$ for two different muon normalizations 
($10^{20}$and $10^{21}$ muon decays of each polarity). For each pair of values 
$(\Delta m^2_{32}, \ \tetonethree)$, the fit was performed leaving
$\theta_{23}$ free. 

In the obtaining previous plot, we 
did not on purpose apply strong background cuts, since we believe that
very high rejection powers obtained on paper may not stand the proof
of real experimental conditions, with non-gaussian behaviours, tails
of distributions etc.\par
To also give the maximum of sensitivity that can be obtained in 
principle, we also illustrated in figure~\ref{fig:sensi} the effect 
that a background free environment would have in the expected sensitivity. 
At 90$\%$ C.L., we obtain 
$\sin^2 2\tetonethree<10^{-3} - 10^{-4}$ depending on the number 
of muon decays. This represents two orders 
of magnitude improvement with respect to quoted sensitivities at CNGS.
In case we consider $10^{21}$ muon decays and backgrounds are reduced
to a negligible level, the obtained sensitivity is consistent with the 
one quoted in~\cite{cervera}.\par
For a background-free environment gives the best sensitivity,
the amount of information added by other event classes (i.e. the
electrons) is negligible.

\subsubsection{Determination of $\theta_{13}$}

The measurement of $\theta_{13}$ can profit from long baselines, since
matter effects will enhance the oscillation signal. In presence of
backgrounds, it is more favorable to enhance neutrinos signal even at
the cost of the suppression of the anti-neutrino oscillations.

Table~\ref{tab:allmeas} summarizes the expected precision on the
measurement of the oscillation parameters.

In this case, since for the chosen value of $sin^2 2\theta_{13}=0.05$
the number of signal events is quite large, there is not any more
a strong need for a background-free environment. Therefore, the
inclusion of other event classes, like the electrons, can help to
constrain the oscillation parameters. 

\begin{table}
\begin{center}
\begin{tabular}{|l|c|c|c|c|}
\hline
\multicolumn{5}{|c|}{Three-family mixing}\\
\multicolumn{5}{|c|}{$\sin^2\theta_{23} = 0.5$, $\sin^22\theta_{13} = 0.05$}\\
\hline
&\multicolumn{2}{|c|}{All classes}&\multicolumn{2}{c|}{Only muons}\\ 
&\multicolumn{2}{|c|}{$\chi^2_{all}$}&\multicolumn{2}{c|}{$\chi^2_{rs\mu}+\chi^2_{ws\mu}$}\\ 
\hline
 & L=2900 km & L=7400 km & L=2900 km & L=7400 km \\ \hline\hline
\multicolumn{5}{|c|}{$\Delta m^2_{32}=3.5 \times 10^{-3}$ eV$^2$, 
} \\ \hline
$\delta(\Delta m^2_{32})$ & $1.4\%$ & $0.9\%$ & 1.4\% & 0.9\%\\
$\delta(\sin^2 \theta_{23})$  & $14\%$ & $8\%$ & 16\% & 9\%\\ 
$\delta(\sin^2 2\theta_{13})$  & $15\%$ & $10\%$ &17\% & 15\% \\ \hline\hline

\multicolumn{5}{|c|}{$\Delta m^2_{32}=5 \times 10^{-3}$ eV$^2$} \\ \hline
$\delta(\Delta m^2_{32})$ & $0.4\%$ & $0.8\%$ & 0.4\% & 0.8\% \\
$\delta(\sin^2 \theta_{23})$  & $11\%$ & $8\%$ & 10\% & 12\%\\ 
$\delta(\sin^2 2\theta_{13})$  & $11\%$ & $9\%$ & 14\% & 16\%\\ \hline\hline

\multicolumn{5}{|c|}{$\Delta m^2_{32}=7 \times 10^{-3}$ eV$^2$} \\ \hline
$\delta(\Delta m^2_{32})$ & $0.4\%$ & $0.6\%$ & 0.4\% & 0.6\% \\
$\delta(\sin^2 \theta_{23})$  & $7\%$ & $8\%$ & 8\% & 18\%\\ 
$\delta(\sin^2 2\theta_{13})$  & $8\%$ & $6\%$ & 9\% & 20\%\\ \hline
\end{tabular}
\caption{Precision on the
measurement of the oscillation parameters.}
\label{tab:allmeas}
\end{center}
\end{table}

\subsection{Sensitivity to $\theta_{13}$ with quasi-elastic events}

\begin{table}
\begin{center}
\begin{tabular}{|c|c|c|c|c|c|c|}
\hline
\multicolumn{7}{|c|}{$\nue$ appearance search with quasi-elastic}\\
\multicolumn{7}{|c|}{Electron Class: Events for $10^{21} \ \mu^-$ decays} 
\\ \hline\hline
 & \multicolumn{2}{|c|}{$\bar{\nu}_e$ CC} & \multicolumn{2}{|c|}{$\nu_\mu\to\nu_e$ CC} & 
\multicolumn{2}{|c|}{$\nu_\mu\to\nu_\tau$ CC, $\tau\to e$} \\
\cline{2-7}
Baseline & Total & Elastic & Total & Elastic & Total & Elastic \\\hline
$L = 732$ km & 860000 & 43000 & 2090 & 84 & 3990 & 110 \\ 
$L = 2900$ km & 54300 & 2700 & 1720 & 70 & 3300 & 90 \\ 
$L = 7400$ km & 8300 & 410 & 960 & 40 & 1450  & 40 \\\hline 
\end{tabular}
\caption{Expected number of electron type events for $10^{21} \ \mu^-$
decays. 
The three contributions to the total number of electron events are 
shown separately. Rates have been computed assuming oscillations with 
$\Delta m^2_{32}= 3.5\times 10^{-3}$ eV$^2$, $\sin^2\theta_{23} = 0.5$
and $\sin^22\theta_{13} = 0.05$.}
\label{tab:qerate}
\end{center}
\end{table}

An exclusive way
of detecting the effects of a non vanishing 
$\tetonethree$ is through the appearance of wrong sign
electrons. Although, with the assumed detector configuration, 
there is no ability to directly measure the charge of
the leading electrons, there is the possibility of disentangling final 
state electrons from positrons through the use of quasi-elastic
events. We look for electron-proton final states. 
For example, in the target of
ICANOE, a proton can be resolved if its kinetic energy is 
larger than about 100~MeV corresponding to a range of more than 2~cm.

Table~\ref{tab:qerate}
shows the expected rates contributing to the electron class 
before any cut is applied. The background is twofold: 
(a) quasi-elastic $\nu_\tau$ CC events followed by $\tau\to e$, 
(b) $\bar{\nu}_e$ CC with an extra proton from 
nuclear origin. 

This last
background can be estimated from data themselves studying the 
reaction $\bar{\nu}_\mu p\to \mu^+ n$. Demanding a back to back
electron-proton event topology with a proton kinetic energy in excess
of 100 MeV, we estimate the expected background to be less than one
event for an overall signal efficiency of $50\%$. 

The quasi-elastic channel
provides, in a ``background free'' environment, 20 to 40 gold-plated 
events depending on the selected baseline. It is very clean channel
but is limited by statistics.

\subsection{Fit of the average Earth density parameter}

\begin{figure}[tb]
\begin{center}
\begin{minipage}{7.5cm}
\epsfig{file=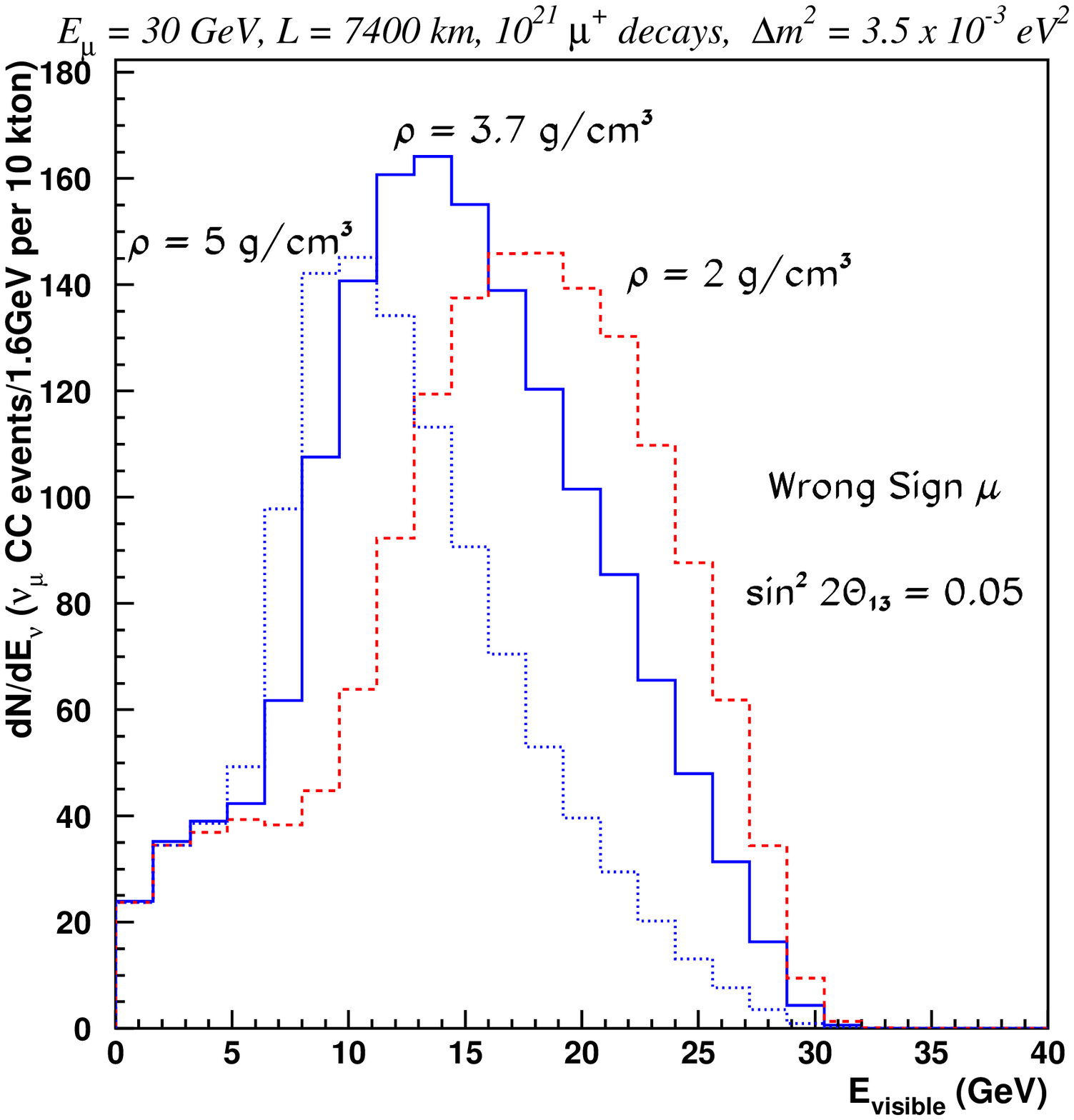,width=7.cm}
\caption{Variation of the MSW resonance peak for wrong sign muons as a function 
of Earth's density. The plot is normalized to $10^{21} \ \mu^+$ decays.}
\label{fig:densipeak}
\end{minipage}
\begin{minipage}{7.5cm}
\epsfig{file=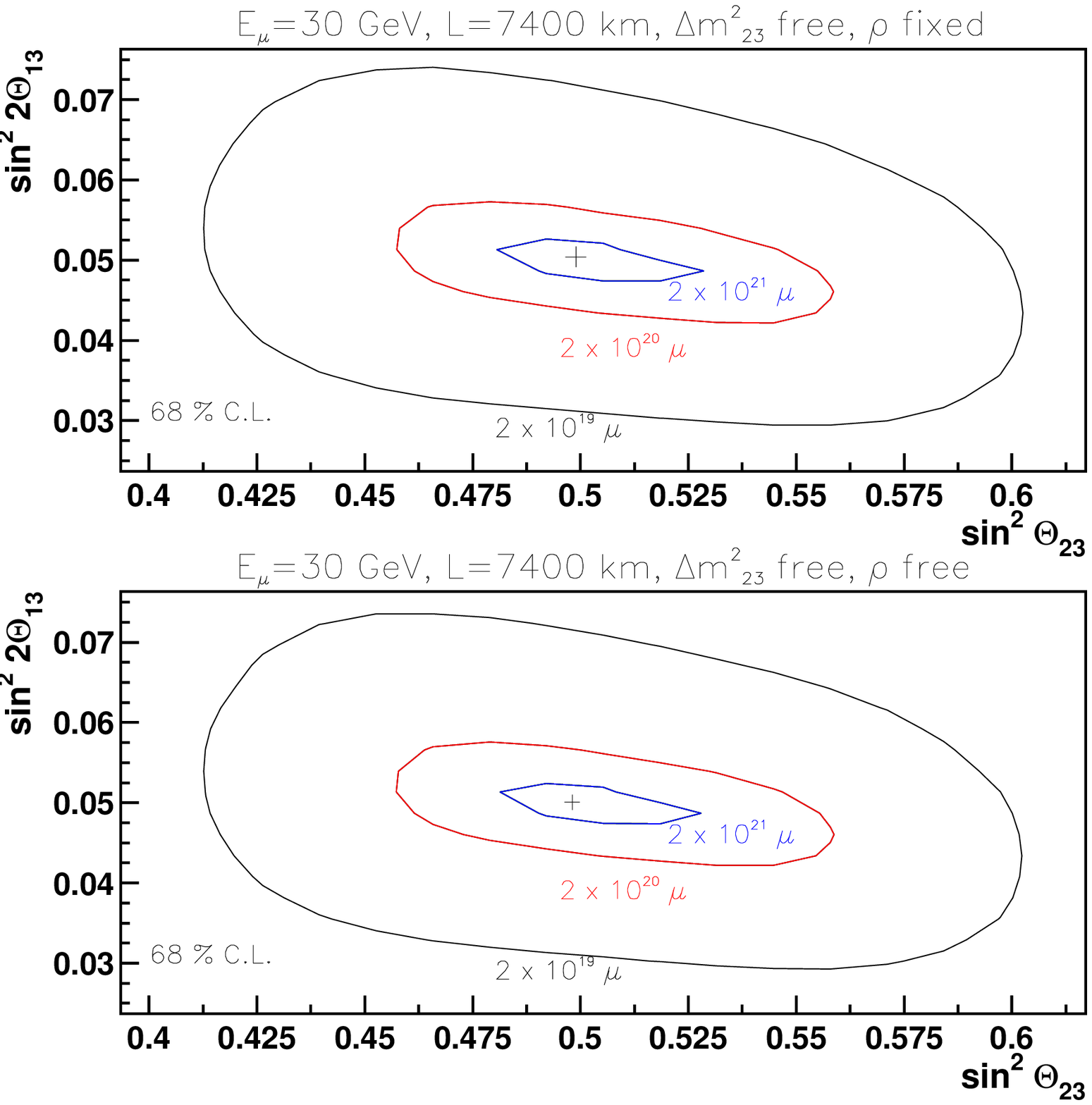,width=7.cm}
\caption{68\%C.L. two-dimensional contours for $\sin^2 2\theta_{13}$
and $\sin^2 \theta_{23}$:
influence of $\rho$ in the determination of the mixing angles 
for three different muon normalizations and $L=7400$ km. In the upper
plot $\rho$ is fixed during the fit, while in the lower one is taken 
as a free parameter.}
\label{fig:rhoinflu}
\end{minipage}
\end{center}
\end{figure}

In matter, the
amplitude of neutrino oscillation goes through a maximum for an energy 
given by equation~\ref{eq:resoene}. 
Since $\tetonethree$ is small, the MSW
resonance peak is only a function of $\rho$ and $\Delta m^2_{32}$, 
This can be seen in Figure~\ref{fig:densipeak}. 
Since the mass
difference is constrained by the disappearance of right-sign muons, 
$\rho$ is well-determined by the energy distribution of wrong sign muons.


We extract the density from the fit, leaving it as a
free parameter, as well as $\Delta m^2_{32}$, $\theta_{23}$ and
$\tetonethree$. The precision on the determination of $\rho$ depends on
the baseline, as shown in table~\ref{tab:dens}, obtained considering
$2\times10^{21}$ muon decays. For the longest baseline, where matter
effect are large, a precision as good as 2\% can be obtained.

\begin{table}
\begin{center}
\begin{tabular}{|c|c|c|}\hline
Distance (km)&Density (g/cm$^3$)&Relative error (\%)\\ \hline
732&2.8&18\\
2900&3.2&10\\
7400&3.7&2\\ \hline
\end{tabular}
\caption{Precision on the determination of $\rho$, from a global fit where also
$\Delta m^2_{32}$, $\theta_{23}$ and $\tetonethree$ are left as free 
parameters. This result has been obtained for $2\times10^{21}$ muon decays.}
\label{tab:dens}
\end{center}
\end{table}

The influence of $\rho$ in our fits is addressed in
figure~\ref{fig:rhoinflu}. We see that for $L=7400$ km and three 
different muon normalizations, the fact that $\rho$ is either considered as a
free parameter or fixed during the fit does not influence the accuracy 
in the determination of the mixing angles.

\section{Over-constraining the oscillation parameters}
\label{sec:taus}

The information about the oscillation parameter is redundantly available
in the visible energy distributions of the various event classes.
This allows us to address the question of the consistency between the
different observed oscillations processes.

Given the high statistical accuracy of the measurement, the
consistency can be tested with good accuracy. 
In the three active neutrino mixing
scheme, this implies that $\Sigma_{y=e,\mu,\tau} P(\nu_x\to\nu_y)$ should 
be equal to one for $x=e,\mu, \tau$ and the same holds for anti-neutrinos. 

Other models can predict different values (i.e., oscillations to sterile 
neutrinos exist, the sum would be smaller than one).\par
Let us concentrate on the oscillations into $\tau$ neutrinos. In the case
of negative muons in the ring, they can be originated from $\nu_\mu\to\nu_\tau$
or $\bar{\nu}_e\to\bar{\nu}\tau$ oscillations. The latter case is particularly
interesting, since coupling a neutrino factory with a detector with $\tau$
identification capabilities is probably the only way to identify and measure
such a process. The $\bar{\nu}_e\to\bar{\nu}\tau$ can be revealed
experimentally from the presence of $\tau$
candidates in the wrong-sign muon sample, due to the opposite helicity
of electron and muon neutrinos in the beam.\par
To have a quantitative estimation of the consistency check of the various
oscillation modes to $\tau$ neutrino, we assign two global normalization 
factors, $\alpha$ and $\beta$ to the oscillation probabilities 
$P(\nu_\mu\to\nu_\tau)$ and $P(\nu_e\to\nu_\tau)$. If no new phenomena occur, 
these parameters should be exactly one. From a global fit to the 
visible energy distributions it is possible to extract the values of these
parameters, and the precision obtainable on their measurement.

\begin{table}[tb]
\begin{center}
\begin{tabular}{|c|c|c|c|c|}
\hline
\multicolumn{5}{|c|}{Appearance/disappearance test}\\
\hline
& & & &  \\
Baseline & $\Delta m^2_{32}$ ($\times 10^{-3}$ eV$^2$) & $10^{20}\mu^\pm$ &
$10^{21}\mu^\pm$ & $10^{22}\mu^\pm$ \\
& & & &  \\
\hline
\multicolumn{5}{|c|}{Precision on $\alpha \Rightarrow \alpha\times P(\nu_{\mu}\to\nu_\tau)$} 
\\ \hline
 & 3.5 & $5.5\%$ & $2\%$ & $0.6\%$\\
7400 km & 5 & $6\%$ & $2\%$ & $0.6\%$\\
 & 7 & $11\%$ & $3\%$ & $1\%$\\\hline
 & 3.5 & $4\%$ & $2\%$ & $0.6\%$\\
2900 km & 5 & $3\%$ & $1\%$ & $0.4\%$\\
 & 7 & $2.5\%$ & $1\%$ & $0.4\%$\\\hline\hline
\multicolumn{5}{|c|}{Precision on $\beta \Rightarrow 
\beta\times P(\nu_e\to\nu_\tau)$} \\ \hline
 & 3.5 & $60\%$ & $20\%$ & $7\%$\\
7400 km & 5 & $35\%$ & $10\%$ & $5\%$ \\
 & 7 & $25\%$ & $7\%$ & $2\%$\\\hline
 & 3.5 & $75\%$ & $25\%$ & $9\%$\\
2900 km & 5 & $25\%$ & $15\%$ & $5\%$\\
 & 7 & $30\%$ & $10\%$ & $4\%$\\\hline
\end{tabular}
\caption{Precision on the determination of the parameters $\alpha$ 
and $\beta$ that quantify, respectively, the amount of
$\nu_\mu\to \nu_\tau$ and $\nu_e\to \nu_\tau$ present in the data. In the three
neutrino framework, the reference values are $\alpha,\ \beta = 1$.}
\label{tab:unitarity}
\end{center}
\end{table}

To select $\tau$ events from the background, still retaining a high efficiency
on the signal, we apply a set of loose kinematic cuts (see 
table~\ref{tab:tausearch}). 
 
Background levels of the order of one event can be reached applying tighter 
cuts, but in these cases the statistics is too small and the results obtained
are slightly worse.

\begin{figure}
\begin{center}
\epsfig{file=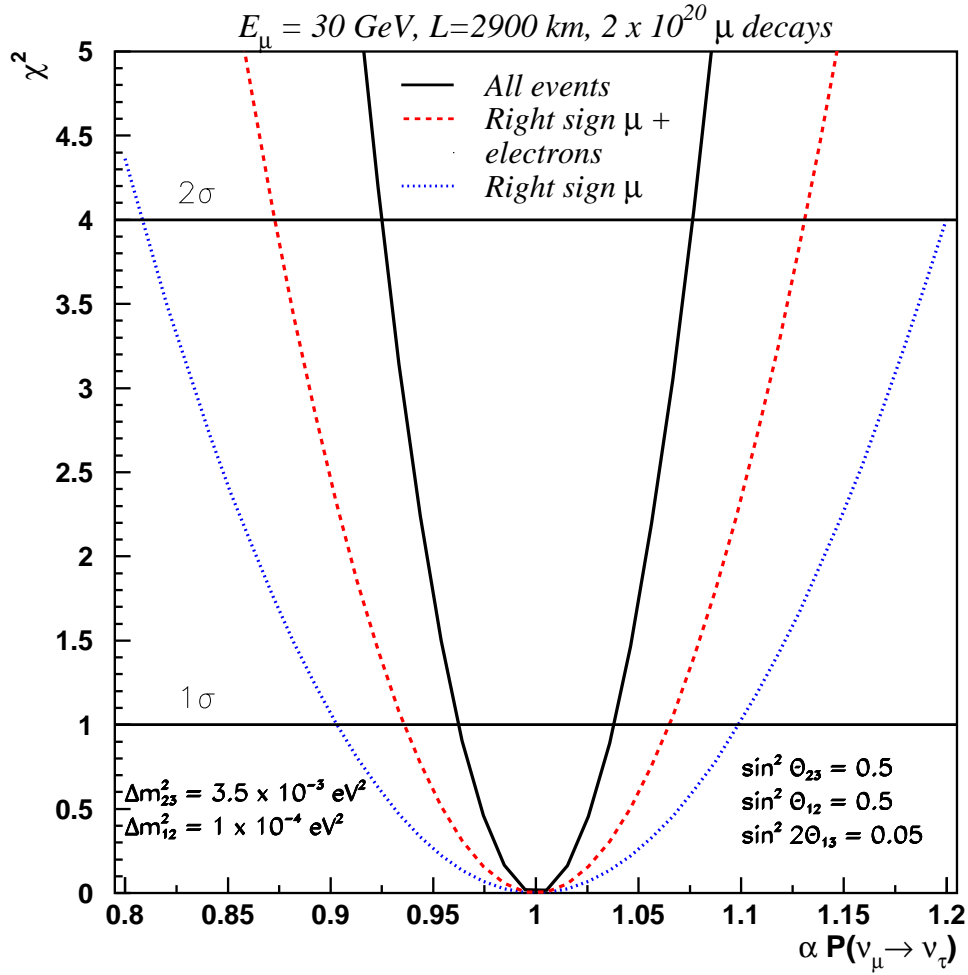,width=14.cm}
\caption{Precision on the determination of the $\alpha$ parameter when 
the fit is performed using: all event classes(solid line), right-sign muons 
+ electrons(dashed line) and right-sign muons only(dotted line). }
\label{fig:taukin}
\end{center}
\end{figure}

Since the $\tau$ lepton decays into muons, electrons and hadrons, we
expect that a fit to all event classes would result in a remarkable 
improvement on the precision for the $\alpha$ parameter. 
Figure~\ref{fig:taukin} shows how $\alpha$ determination improves as
the different event classes are included in the fit.

Table~\ref{tab:unitarity} shows the
expected precisions in the determination of $\alpha$ and $\beta$. 
We observe that for $10^{20}$ decays, $\alpha$ is better 
determined (accuracy around $5\%$) for $L=2900$km, 
however for $10^{21}$ muons, the accuracy is about
$1\%$ regardless of the baseline and the mass difference and therefore 
$\nu_\mu$ oscillations into a sterile neutrino can be largely ruled out. 

The accuracy on $\beta$, and therefore the first experimental 
evidence for $\nu_e\to\nu_\tau$ is much worst, given the smaller
statistics available, since this oscillation probability is smaller
than the corresponding $\nu_\mu\to\nu_\tau$ by a factor
$\sin^2 2\theta_{13}$, taken to be in this case 0.05.
We observe that somewhat better determination exists at $L=7400$km 
since this oscillation mode is largely influenced by matter effects
given its dependence on $\theta_{13}$. 
We conclude that a precision at the level of a few per cent 
on the observation of $\nu_e\to\nu_\tau$ oscillations would require 
$O(10^{22})$ useful muon decays. 

\section{Results for the general three-family scenario and CP-violation}
\label{sec:cp}
Let us consider now a more complex scenario. In this case, the value of the
mass difference $\Delta m^2_{12}$ is not any more negligible, and is actually
assumed to be $10^{-4} eV^2$, one of the highest values compatible with the
large mixing angle MSW solution for solar neutrinos \cite{eprof}.

The oscillation does not depend any more on only three parameters, but all
four independent angles of the mixing matrix and the two mass differences
become important.

In the most general case, the phase $\delta$ can be
different from zero, producing a complex mixing matrix, and thus generating 
CP violation.

We recall that for neutrinos propagating in vacuum, the oscillation
probability after a distance $L$ for neutrinos can be expressed as: 
\begin{eqnarray}\label{eq:oscill2}
P(\nu_\alpha\ra\nu_\beta;E,L)& = & P_{CP \
even}(\alpha,\beta;E,L)
+ P_{CP \ odd}(\alpha,\beta;E,L)
\end{eqnarray}
and for antineutrinos : 
\begin{eqnarray}\label{eq:oscill3}
P(\bar{\nu}_\alpha\ra\bar{\nu}_\beta;E,L)& = & P_{CP \
even}(\alpha,\beta;E,L)
- P_{CP \ odd}(\alpha,\beta;E,L)
\end{eqnarray}

As an illustration, the probability for $\nu_\mu$ to $\nu_e$
conversion in vacuum, assuming three family mixing and CP-violation, is
given by:
\begin{eqnarray}
 P_{CP\ odd}(\mu,e)& = &
2{\cos (\theta^m_{13})}^2
\sin\delta^m\sin (2\theta^m_{12})\sin 
(\theta^m_{13})\sin (2\theta^m_{23})\times \\ \nonumber
& & \sin (\frac{\Delta M_{12}^2 L}{4E})
  \sin (\frac{\Delta M_{13}^2  L}{4E})
  \sin (\frac{\Delta M_{23}^2  L}{4E})
\end{eqnarray}
where the $CP$-phase in matter is:
\begin{equation}
e^{-i\delta_m}  = 
\frac{(G^2e^{-i\delta}-F^2e^{i\delta})s_{23}c_{23}+
GF(c^2_{23}-s^2_{23})}{\sqrt{(G^2s^2_{23}+F^2c^2_{23}+2GFc_{23}s_{23}c_\delta)
(G^2c^2_{23}+F^2s^2_{23}-2GFc_{23}s_{23}c_\delta)}} 
\end{equation}
Here, $G$ and $F$ are given in the Appendix.

In case of degenerate mass differences, $|\Delta m^2_{21}|\approx |\Delta m^2_{32}|$,
the $\delta$ phase is significantly modified from its
original value in vacuum for $\delta > \pi/4$ radians, while in the case 
$|\Delta m^2_{32}| > |\Delta m^2_{21}|$ the CP phase is almost
unaffected by the presence of matter.

In addition, neutrino propagation in a dense
medium makes more difficult to experimentally extract a genuine CP violation
signal, since the asymmetrical behavior of matter, with respect to 
neutrinos and anti-neutrinos, induces fake CP violation effects.

\subsection{$CP$ statistical significance}
\begin{figure}
\begin{center}
\epsfig{file=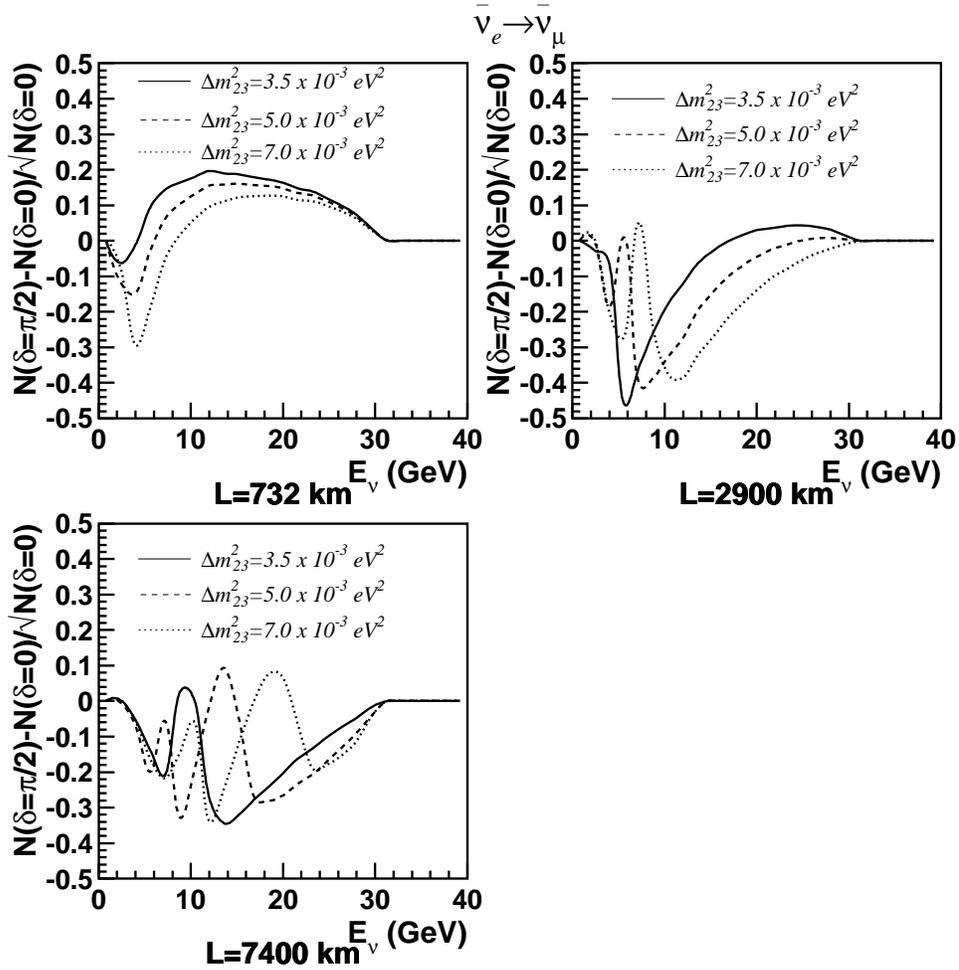,width=14.cm}
\caption{Statistical significance of CP violation effects for three 
possible $\Delta m^2_{32}$ values mass at the three considered baselines.
$N(\delta=\pi/2)$ is the number of expected events assuming maximal CP 
violation in vacuum and $N(\delta=0)$ corresponds to the number of expected 
events in absence of CP violation, for $2\times 10^{21}$ muons}
\label{fig:cpsignificance}
\end{center}
\end{figure}

To evaluate the sensitivity to this kind of measurement as a function of the
oscillation parameters and of the selected baseline, we compare the case
where CP violation is maximal ($\delta=\pi/2$) and the case of no CP
violation ($\delta=0$). Figure~\ref{fig:cpsignificance} shows the sensitivity 
\begin{equation}
S_{CP}(E_i)\equiv\frac{N(\delta=\pi/2,E_i)-N(\delta=0,E_i)}{\sqrt{N(\delta=0,E_i)}}
\end{equation}
i.e. the difference between the two extreme cases, divided by the
statistical error (being N the number of events in each energy bin, for
a given value of $\delta$). We can see that for the baseline of 732 km,
this quantity is positive for almost the full energy range, so there is no
real shape variation in the spectrum, and the CP effect is similar to what
would be obtained with a change in the angle $\theta_{13}$. On the other
hand, for larger baselines this curve crosses the zero, and the CP
violation produces a visible deformation of the energy spectrum.

\subsection{Fitting of the $\delta$ parameter}

To evaluate the precision reachable on the measurement of these three
oscillations parameters, we perform fits assuming the following
reference values: 
\begin{eqnarray}
\Delta m^2_{32} & = & 3.5,5, 7 \times 10^{-3} \ {\rm eV}^2\nonumber \\
\Delta m^2_{12} & = & 1 \times 10^{-4} \ {\rm eV}^2\nonumber \\
\sin^2 \tet23 & = & 0.5  \hspace*{3.2cm}\tet23 = 45^o \nonumber \\
\sin^2 2\tetonethree & = & 0.05 \hspace*{2.9cm}\tetonethree = 6.5^o
\nonumber \\
\sin^2 \theta_{12} & = & 0.5 \nonumber \\
\delta \geq 0  
\end{eqnarray}

\begin{figure}
\begin{center}
\epsfig{file=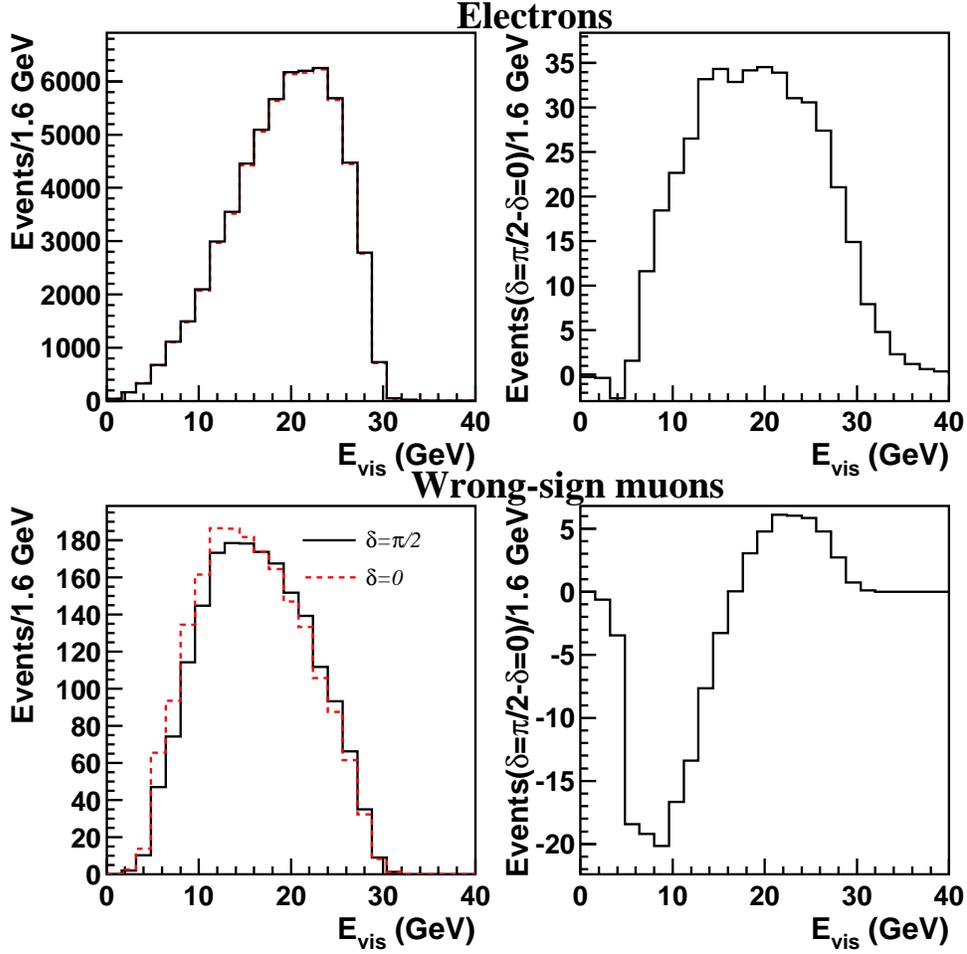,width=14.cm}
\caption{Energy spectra and differences between no CP violation ($\delta$=0)
or maximal CP violation ($\delta = \pi/2$). The two upper plots refer to the
electron class, the lower ones to the wrong-sign muon class.
The plots on the left show the energy spectra for the
two cases, the ones on the right the different in number of events as a 
function of the energy. The helicity of muons circulating in the ring
has been chosen in such a way to enhance the effect, i.e. negative
muons have been used for the upper plots, and positive ones for the
lower plots.}
\label{fig:cpvar}
\end{center}
\end{figure}

Figure~\ref{fig:cpvar} shows for the baseline L=2900 km, the 
expected visible energy spectra for electron and wrong sing muon
class for the two cases $\delta=0$ and $\delta=\pi/2$, and their
difference in terms of number of events. As expected, in absolute 
values, the CP violation affects the two classes in a similar way,
but the effect is much more visible and significant for the
wrong-sign muons, since the total number of events is smaller.

Figure~\ref{fig:cpvar} is 
normalized to $10^{21} \ \mu^+$ decays, since this is the minimum
amount of decays required to produce a statistically significant 
observation of CP violation.

\begin{figure}
\begin{center}
\epsfig{file=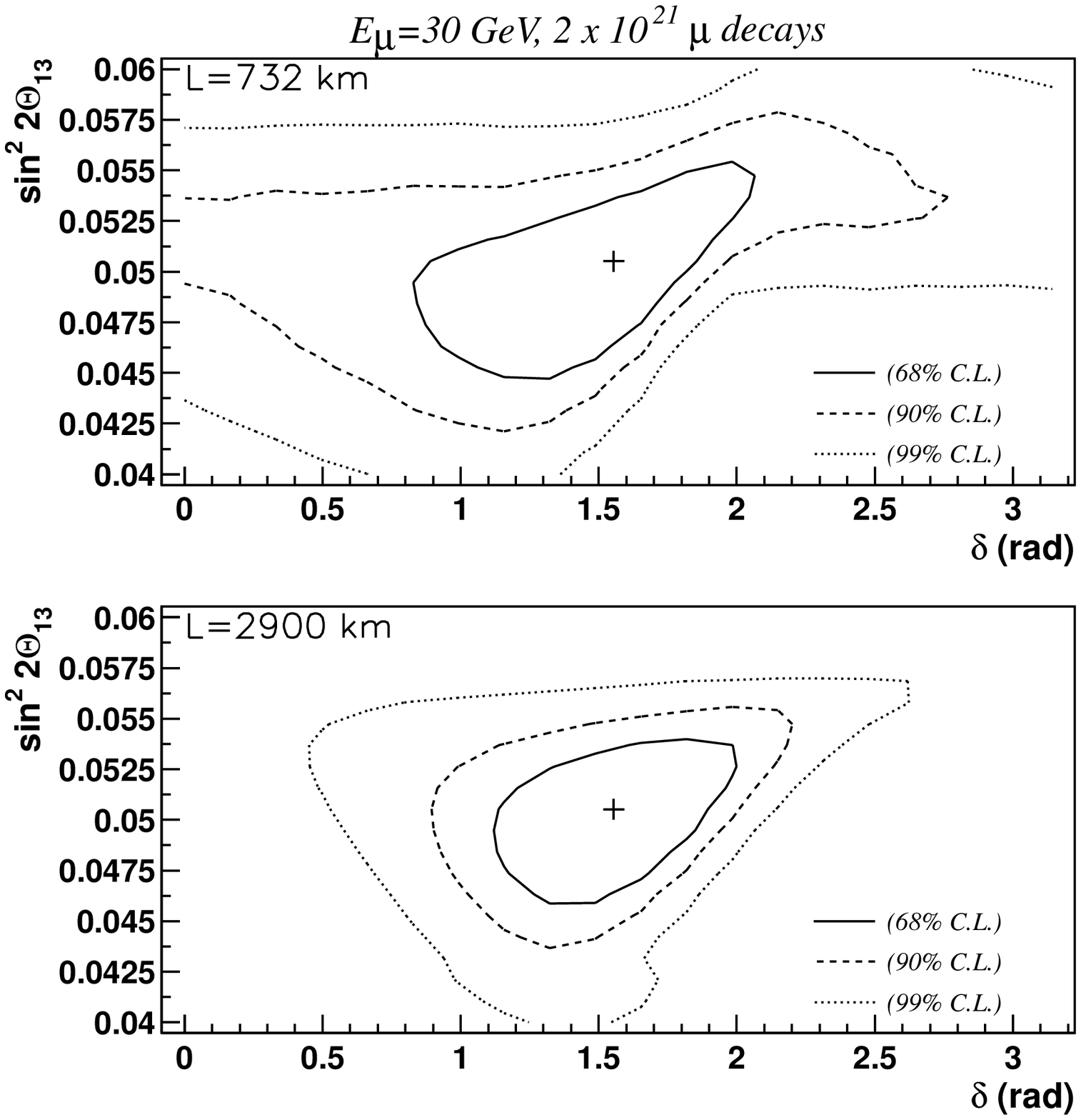,width=14.cm}
\caption{Correlation between $\theta_{13}$ and CP phase $\delta$ for
two different baselines and $2 \times 10^{21}$ decays.}
\label{fig:cpvst13}
\end{center}
\end{figure}

For the largest baseline $L=7400$ km, CP violation effects are almost
undetectable even in case $\delta = \pi/2$ since, wrong sing muon
appearance is largely dominated by matter effects. At $L=732$ km, 
the effect of a non-vanishing $\delta$ is more striking thanks to 
the higher event rates and the smaller neutrino path in
matter. However, as we formerly pointed out, 
this effect is similar to the one produced by a
smaller value of $\theta_{13}$ and therefore, both effects cannot be
disentangled with a single measurement at this baseline as shown in 
figure~\ref{fig:cpvst13}, where the correlations between $\delta$
and $\theta_{13}$ prevent a precise determination of any of them.
Nonetheless, a measurement at $L=7400$ km where CP violation effects
are negligible and the most accurate determination of $\theta_{13}$
exits, can be combined with data collected at $L=732$ km to produce a
precise determination of $\delta$. 

Another possibility to unveil the existence of CP violation is to
perform a single measurement at $L=2900$ km. As shown in 
figure~\ref{fig:cpvar}, at this distance the effect of $\delta \neq 0$ 
is twofold: not only the event rate is modified but also the spectral
shape. This last effect cannot be produced by a change on
$\theta_{13}$. Figure~\ref{fig:cpvst13} shows that at this distance
the correlation between $\delta$ and $\theta_{13}$ has diminished and
therefore a better determination of the parameters can be achieved
with a single measurement.

\begin{figure}
\begin{center}
\epsfig{file=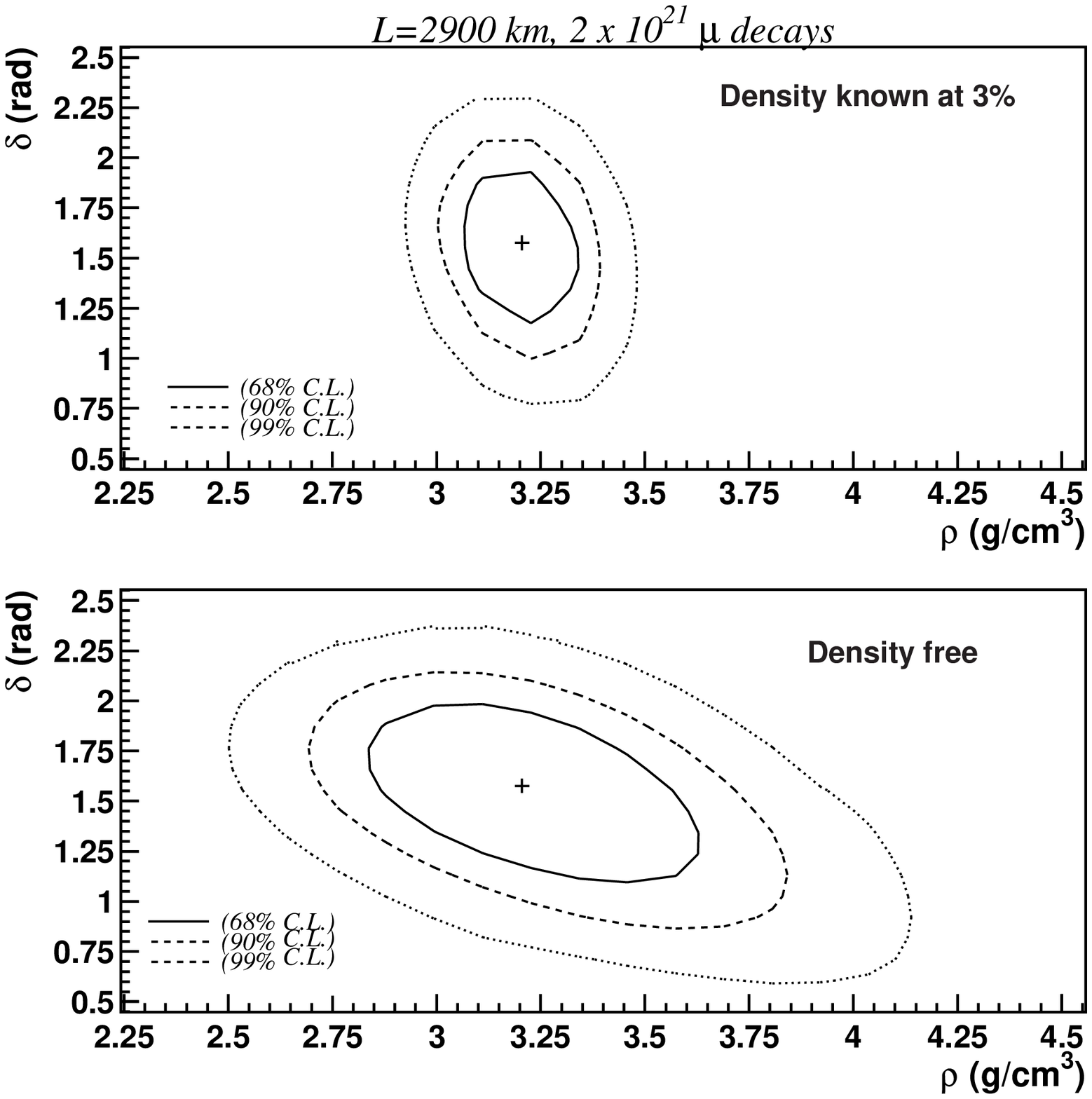,width=14.cm}
\caption{Correlation between the average matter density $\rho$
and the CP phase $\delta$ for L=2900 km. In the lower plot we 
leave Earth's mean density as a free parameter in the fit. In the 
upper plot we assume that density is known within $3\%$. 
We see that the presence of matter does not spoil the possibility 
of performing a measurement of $\delta$.}
\label{fig:dvsrho}
\end{center}
\end{figure}

For a baseline of $L=2900$ km, matter effects are not so strong as for the
longest baseline. The possibility of measuring the CP-violating phase 
$\delta$ is not spoiled by the fake asymmetries due to the matter 
interactions, but both effects, even if correlated, can be measured 
at the same time. This is shown in figure~\ref{fig:dvsrho}, where 
the result of a simultaneous fit to the average matter density 
$\rho$ and the phase $\delta$ is presented for the two cases where 
$\rho$ is left as a free parameter or it is known beforehand with 
a $3\%$ accuracy.

\begin{figure}
\begin{center}
\epsfig{file=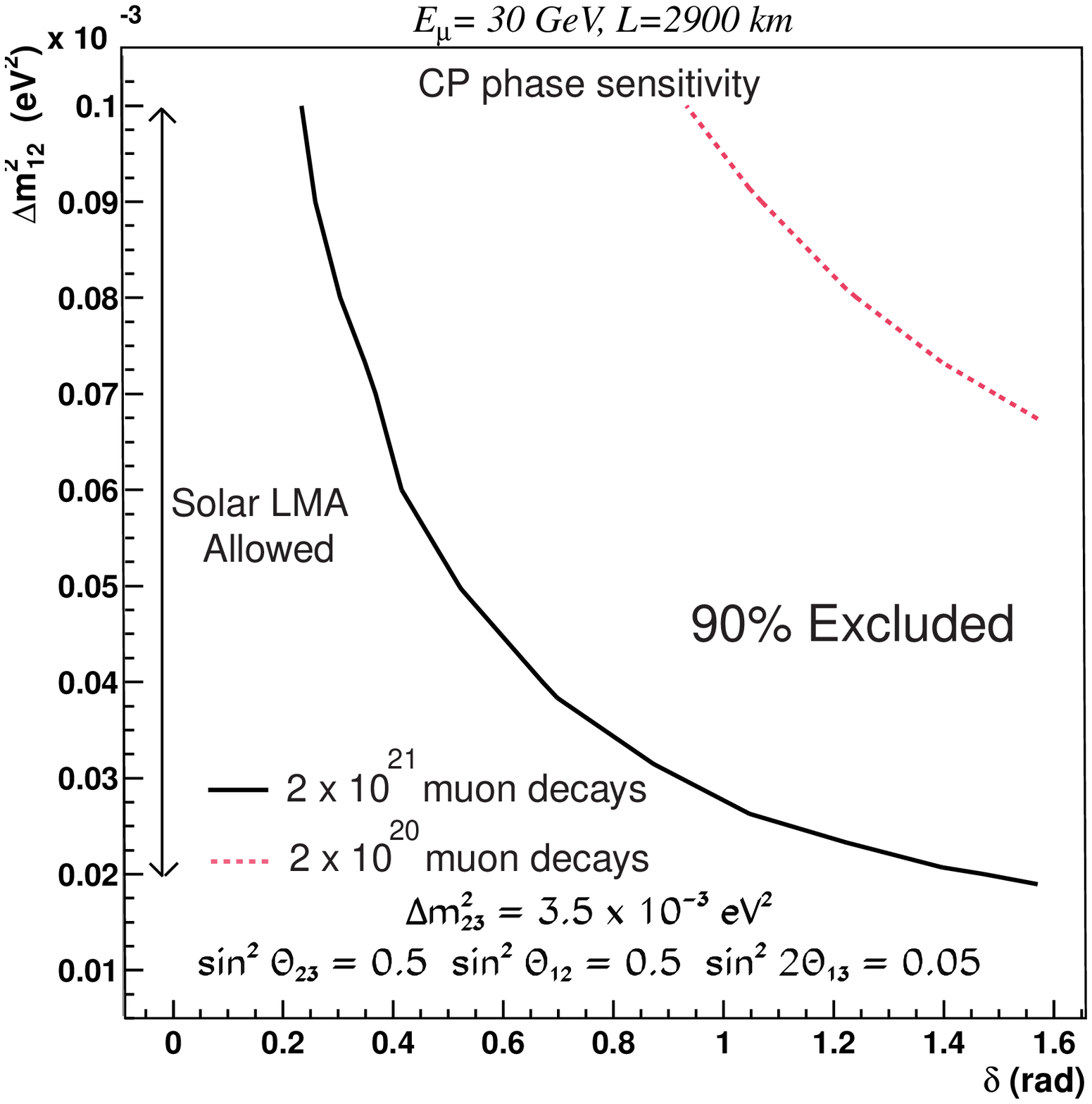,width=14.cm}
\caption{90\%C.L. sensitivity on the CP phase $\delta$ as a function of
$\Delta m^2_{12}$ for two different normalizations: solid (dashed)
line corresponds to $10^{21} \ (10^{20})$ muons decays of each polarity.}
\label{fig:cpsensi}
\end{center}
\end{figure}

\begin{figure}
\begin{center}
\epsfig{file=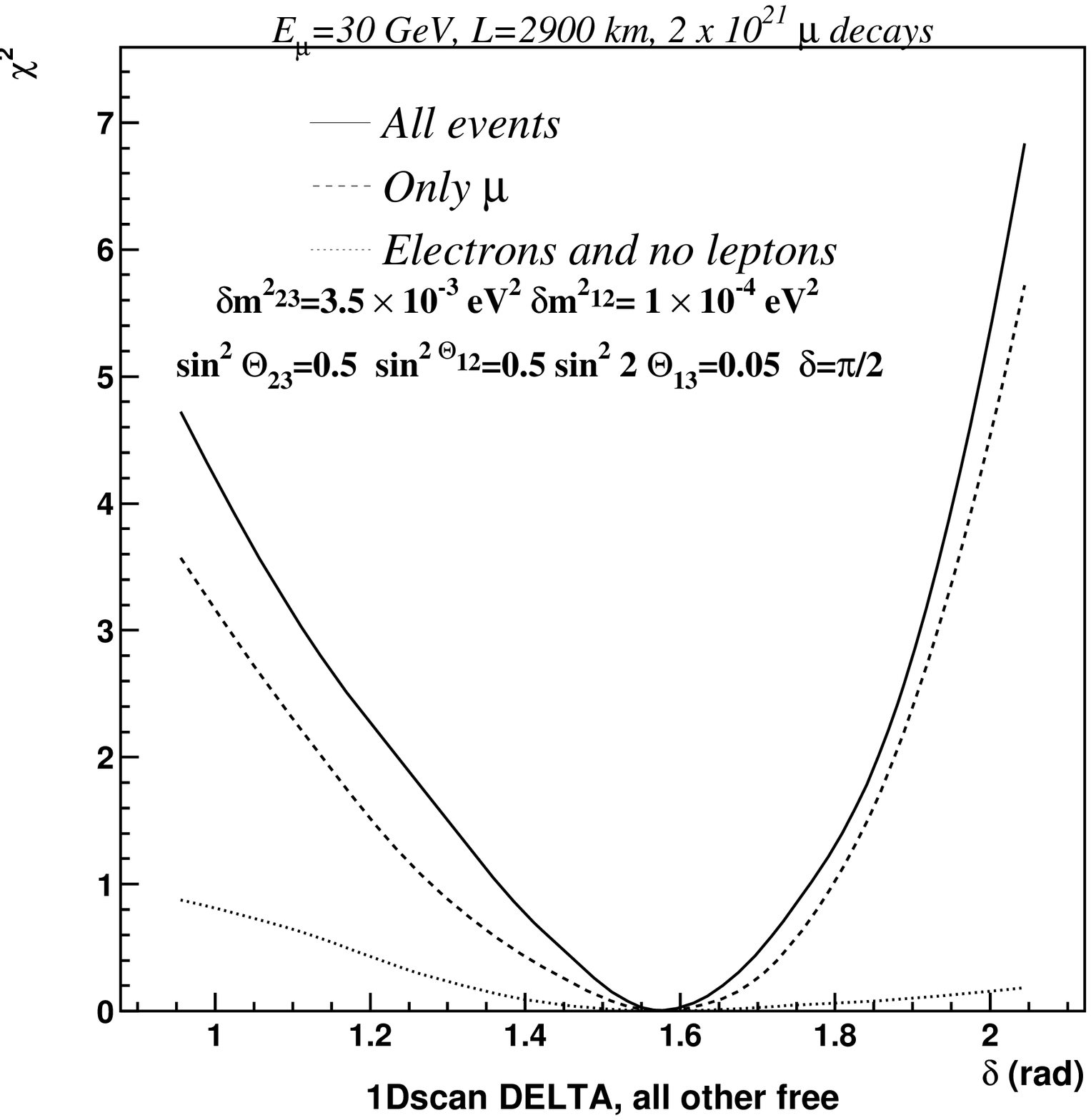,width=14.cm}
\caption{Measurement of the CP violating phase $\delta$ for
$2\times10^{21} \ \mu$ decays and $L=2900$ km. The two mass
differences and the three mixing angles have been left free during the 
fit. The reference value is $\delta = \pi/2$. The three curves are obtained
using respectively in the fit all classes, only the muon classes, or only 
electrons and NC-like events.}
\label{fig:cpmeas}
\end{center}
\end{figure}

In case no CP violation is observed, the allowed $\delta$ values at 
90$\%$ C.L. as a function of $\Delta m^2_{12}$ are shown in
figure~\ref{fig:cpsensi} for two different muon normalizations. 
On the other hand, if a significant effect is detected, 
the precision achievable on the measurement of the CP phase $\delta$ 
is shown in figure~\ref{fig:cpmeas}. We 
fit $\delta$ using all event classes, leaving the five parameters
governing the oscillation free. Assuming a reference value of $90^o$
for $\delta$ and $10^{21}$ muon decays of each polarity, 
we get: $90 \pm 15 ^o$. Therefore a precision around
20$\%$ is expected. Finally, we note that the change on the 
expected precision for $\theta_{13}$ is negligible when 
$\Delta m^2_{12}$ and $\theta_{12}$ are included in the fit.

\subsection{Use of quasi-elastic events}
Quasi-elastic events can also be useful to spot the
presence of CP violation.  
Table~\ref{tab:qecp} shows the expected number of QE 
electron events after kinematics cuts (a back to back 
electron-proton final state with proton kinetic energy in excess of 
100 MeV). Three different assumptions for the total 
number of muon decays and the baseline of 2900~km have been
assumed.
In case CP is conserved, we expect 
35 (96) quasi-elastic electron events for $\Delta m^2_{32} = 3.5
(7) \times 10^{-3}$ eV$^2$ and $10^{21}$ muon decays of each
polarity. In case $\delta=\pi/2$, we expect 26 and 85 events for the 
two mass differences considered. The effect is at the one sigma
level. To obtain a statistically conclusive signal for 
CP violation would require more than $10^{21}$ muon decays. 
\begin{table}
\begin{center}
\begin{tabular}{|l|r|c|c|c|}
\hline
\multicolumn{5}{|c|}{CP-violation with quasi-elastic events}\\
\hline
\multicolumn{2}{|c|}{L=2900 km } & $N_{ele}$ & 
$N_{ele}$ & Stat. \\ 
\multicolumn{2}{|c|}{ } & $(\delta=0)$ & $(\delta=\pi/2)$ & significance \\ 
\hline
$\Delta m^2_{32} = 3.5\times 10^{-3} eV^2$ & $10^{21}$ $\mu^\pm$ & 
35 & 26 & $1.5\sigma$ \\
$\sin^2\theta_{23} = 0.5$, $\sin^22\theta_{23} = 0.05$ 
& $5\times10^{21}$ $\mu^\pm$ & 
175 & 130 & $3.4\sigma$ \\
$\Delta m^2_{12} = 10^{-4}$ eV$^2$, $\sin^2\theta_{12} = 0.5$ & $10^{22}$ $\mu^\pm$ & 350 & 260 & $4.8\sigma$ \\\hline\hline
$\Delta m^2_{32} = 7\times 10^{-3}$ eV$^2$ & $10^{21}$ $\mu^\pm$ & 96 & 85 & $1.1\sigma$ \\
$\sin^2\theta_{23} = 0.5$, $\sin^22\theta_{23} = 0.05$ 
& $5\times10^{21}$ $\mu^\pm$ & 
480 & 425 & $2.5\sigma$ \\
$\Delta m^2_{12} = 10^{-4}$ eV$^2$, $\sin^2\theta_{12} = 0.5$ & $10^{22}$ $\mu^\pm$ & 960 & 850 & $3.6\sigma$ \\\hline
\end{tabular}
\caption{Expected QE electron events ($N_{ele}$) at a baseline of 2900 km 
in case there is no CP violation and in case CP violation phase in vacuum is 
maximal. The last column shows, in number of sigmas, the statistical 
significance expected for the QE electron sample in case CP is violated 
in the lepton sector.}
\label{tab:qecp}
\end{center}
\end{table}

\section{Conclusions}

In this paper, we tried to understand in deeper detail the 
capabilities an experiment at the Neutrino Factory, with the aim of
exploiting as much as possible the possibility of studying several 
neutrino transitions at the same time. For this reason, we have 
considered as a baseline detector a large Liquid Argon TPC with
external muon identifier, the most versatile design proposed so far
for large neutrino experiments.\par
Assuming an oscillation scenario favoured by present experimental
results, the leading oscillation would be between the second and
third neutrino family, which are maximally mixed. 
Therefore, a very precise 
determination of the parameters governing this transition, 
$\theta_{23}$ and $\Delta m^2_{23}$ is essential also for the
understanding of all other processes. This is mainly achieved using
the information coming from the $\nu_\mu$ disappearance (right-sign
muon class), provided that the baseline and beam energy are chosen
in such a way that the first oscillation maximum is visible as a dip
in the oscillated spectrum.\par
The maximal sensitivity to $\theta_{13}$ is achieved for very small
background levels, since we are looking in this case for small 
signals; most of the information is coming from the clean
wrong-sign muon class, and from quasi-elastic events.\par
On the other hand,  if its value is not too small, for a 
measurement of $\theta_{13}$, the signal/background ratio could be
not so crucial, and also the other event classes can contribute to 
this measurement.\par
Like for a B-Factory, a $\nu$-Factory should have among its aims
the overconstraining of the oscillation pattern, in order to look for
unexpected new physics effects. This can be achieved in global
fits of the parameters, where the unitarity of the mixing matrix is 
not strictly assumed.
Using a detector able to identify the $\tau$ lepton production via
kinematic means, it is possible to verify the unitarity in 
$\nu_\mu\to\nu_\tau$ and $\nu_e\to\nu_\tau$ transitions. For this 
latter, the possibility of a kinematical $\tau$ identification
for wrong-sign muon events could allow for the first time a clear
identification of this type of oscillations.\par
The study of CP violation in the lepton system is a very
fascinating subject, and probably the most ambitious goal of this
kind of machines. It will only be possible for high beam intensities,
and if the parameters governing the solar neutrino deficit are in
the region usually indicated as large mixing angle MSW solution.
Matter effect can mimic CP violation; however, a multiparameter fit
at the right baseline can allow a simultaneous determination of
matter and CP-violating parameters.\par
Also for CP-violation measurements most of the information would 
come from the wrong-sign muon class, but since in this case 
the electron class would also be affected, the study of these events
(and of the very clean quasi-elastic interactions) can provide 
essential cross-checks for these delicate measurements.

%
%
\appendix
\section{Oscillations in matter} 
\label{sec:oscmat}

In reference~\cite{zaglauer}, the authors compute analytic expressions 
for the mass eigenvalues, mixing angles and CP-violation phase in
matter, assuming the mixing matrix $U$ is parametrized 
``\`a  la CKM''. 
Since several missprints were observed,
we reproduce here the corrected expressions: 

For neutrinos, the mass eigenvalues in matter $M_1, \ M_2$ and $M_3$ are: 
\begin{eqnarray}
M_1^2 & = & m_1^2 +\frac{A}{3} - \frac{1}{3}\sqrt{A^2-3B}S -
\frac{\sqrt{3}}{3}\sqrt{A^2-3B}\sqrt{1-S^2}\\
M_2^2 & = & m_1^2 +\frac{A}{3} - \frac{1}{3}\sqrt{A^2-3B}S +
\frac{\sqrt{3}}{3}\sqrt{A^2-3B}\sqrt{1-S^2}\\
M_3^2 & = & m_1^2 +\frac{A}{3} + \frac{2}{3}\sqrt{A^2-3B}S
\end{eqnarray}
where
\begin{eqnarray}
A & = & \Delta m^2_{21} + \Delta m^2_{31} + D \\
B & = & \Delta m^2_{21} \Delta m^2_{31} + D[\Delta m^2_{31} c^2_{13}
+ \Delta m^2_{21}(c^2_{13}c^2_{12}+s^2_{13})] \\
C & = & D\Delta m^2_{21}\Delta m^2_{31}c^2_{13}c^2_{12} \\
D & = & 2\sqrt{2}G_F N_e E \rm\ for\ neutrinos\\
S & = & \cos\left[\frac{1}{3}\arccos\left(
\frac{2A^3-9AB+27C}{2\sqrt{(A^2-3B)^3}}\right)\right]
\end{eqnarray}
and $c_{ij} = \cos \theta_{ij}$ and $s_{ij} = \sin \theta_{ij}$.
The mixing angles and CP phase in matter are: 
\begin{eqnarray}
\sin^2\theta^m_{12} & = & \frac{-(M_2^4-\alpha M_2^2+\beta)\Delta M^2_{31}}
{\Delta M^2_{32}(M_1^4 -\alpha M_1^2+\beta)-\Delta
M^2_{31}(M^4_2-\alpha M_2^2 + \beta)} \\
\sin^2\theta^m_{13} & = & \frac{M_3^4-\alpha M_3^2+\beta}{\Delta
M^2_{31}\Delta M^2_{32}} \\
\sin^2\theta^m_{23} & = &
\frac{G^2s^2_{23}+F^2c^2_{23}+2GFc_{23}s_{23}c_\delta}{G^2+F^2} \\
e^{-i\delta_m} & = &
\frac{(G^2e^{-i\delta}-F^2e^{i\delta})s_{23}c_{23}+
GF(c^2_{23}-s^2_{23})}{\sqrt{(G^2s^2_{23}+F^2c^2_{23}+2GFc_{23}s_{23}c_\delta)
(G^2c^2_{23}+F^2s^2_{23}-2GFc_{23}s_{23}c_\delta)}} 
\end{eqnarray}
where 
\begin{eqnarray}
\alpha & = & m_3^2 c^2_{13} + m_2^2(c^2_{13}c^2_{12}+s^2_{13}) +
m_1^2(c^2_{13}s^2_{12}+s^2_{13}) \\
\beta & = & m_3^2 c^2_{13} (m_2^2c^2_{12}+m_1^2s^2_{12}) + m_2^2 m_1^2 
s^2_{13} \\
G & = & [\Delta m^2_{31}(M_3^2-m_1^2-\Delta m^2_{21})-\Delta m^2_{21} 
(M_3^2-m_1^2-\Delta m^2_{31})s^2_{12}]c_{13}s_{13} \\
F & = & (M_3^2-m_1^2-\Delta m^2_{31})\Delta m^2_{21}c_{12}s_{12}c_{13}
\end{eqnarray}

For anti-neutrinos, we must replace $D$ by $-D$.

In the case of one-mass scale approximation, the oscillation
probabilities are simply
The oscillation probabilities are: 
\begin{eqnarray}\label{eq:probs}
P(\nue\ra\nue,E,L) & = & 1-\sin^2(2\theta^m_{13})\Delta^2_{32} \\
P(\nue\ra\numu,E,L) & = & \sin^2(2\theta^m_{13})\sin^2(\theta^m_{23})\Delta^2_{32}\nonumber\\
P(\nue\ra\nutau,E,L) & = & \sin^2(2\theta^m_{13})\cos^2(\theta^m_{23})\Delta^2_{32} \nonumber\\
P(\numu\ra\numu,E,L) & = & 1-4\cos^2(\theta^m_{13})\sin^2(\theta^m_{23}) \left[1-\cos^2(\theta^m_{13})\sin^2(\theta^m_{23})\right]\Delta^2_{32}\nonumber\\
P(\numu\ra\nutau,E,L) & = &
\cos^4(\theta^m_{13})\sin^2(2\theta^m_{23})\Delta^2_{32}\nonumber
\end{eqnarray}
where $\Delta^2_{32} =  \sin^2\left((M_3^2-M_2^2) L/4E\right)$. 

%
%

\end{document}